\documentclass[11pt,finalcls,onecolumn]{IEEEtran}
\usepackage{graphicx}
\usepackage{amsmath}
\usepackage{mathrsfs}
\usepackage{amssymb}
\usepackage{float}
\usepackage{url}
\usepackage{verbatim}
\usepackage{dsfont}
\usepackage{enumerate}

\usepackage[bottom=0.8in]{geometry}

\newtheorem{theorem}{Theorem}
\newtheorem{lemma}{Lemma}
\newtheorem{proposition}{Proposition}
\newtheorem{corollary}{Corollary}

\newtheorem{definition}{Definition}
\newtheorem{example}{Example}
\newtheorem{remark}{Remark}

\newcommand{\naturals}{\ensuremath{\mathbb{N}}}

\begin{document}
\title{Projection Theorems for the R\'enyi Divergence on $\alpha$-Convex Sets}

\author{M. Ashok Kumar \qquad Igal Sason
\thanks{
M. Ashok Kumar is with the Indian Institute of Technology, Indore 453552, India
(e-mail: ashokm@iiti.ac.in). The work was partially done while he was a post-doctoral fellow
at the Andrew and Erna Viterbi Faculty of Electrical Engineering, Technion--Israel
Institute of Technology, Haifa 32000, Israel.}
\thanks{
I. Sason is with the Andrew and Erna Viterbi Faculty of Electrical Engineering,
Technion--Israel Institute of Technology, Haifa 32000, Israel (e-mail: sason@ee.technion.ac.il).}
\thanks{
This work has been supported by the Israeli Science Foundation (ISF) under
Grant 12/12. It has been presented in part at the
2016 IEEE International Symposium on Information Theory, Barcelona,
Spain, July~10--15, 2016.}}

\maketitle

\begin{abstract}
This paper studies forward and reverse projections for the
R\'{e}nyi divergence of order $\alpha \in (0, \infty)$ on
$\alpha$-convex sets. The forward projection on such a set
is motivated by some works of Tsallis {\em et al.} in
statistical physics, and the reverse projection is motivated
by robust statistics. In a recent work,
van Erven and Harremo\"es proved a Pythagorean inequality
for R\'{e}nyi divergences on $\alpha$-convex sets under the
assumption that the forward projection exists. Continuing
this study, a sufficient condition for the existence of a forward
projection is proved for probability measures on a general
alphabet. For $\alpha \in (1, \infty)$,
the proof relies on a new Apollonius theorem for the Hellinger
divergence, and for $\alpha \in (0,1)$, the proof relies on the
Banach-Alaoglu theorem from functional analysis. Further projection
results are then obtained in the finite alphabet setting. These
include a projection theorem on a specific $\alpha$-convex set,
which is termed an {\em $\alpha$-linear family}, generalizing a
result by Csisz\'ar to $\alpha \neq 1$.
The solution to this problem yields a parametric family of probability
measures which turns out to be an extension of the exponential
family, and it is termed an {\em $\alpha$-exponential family}.
An orthogonality relationship between the $\alpha$-exponential
and $\alpha$-linear families is established, and it is used to
turn the reverse projection on an $\alpha$-exponential family
into a forward projection on an $\alpha$-linear family. This paper
also proves a convergence result of an iterative
procedure used to calculate the forward projection on
an intersection of a finite number of $\alpha$-linear families.
\end{abstract}


\eject

\section{Introduction}
\label{sec:introduction}
Information projections of relative entropy have been extensively studied due
to their various applications in large deviations theory (e.g., Sanov's
theorem and the conditional limit theorem), maximum likelihood estimation (MLE),
statistical physics, and so on. Some of the pioneering works studying information
projections include Barron \cite{Barron00}, \={C}encov \cite{Cencov82_book},
Chentsov \cite{Chentsov68}, Csisz\'ar \cite{Csiszar75}, \cite{Csiszar84},
Csisz\'ar and Mat\'{u}\u{s} \cite{CsiszarM03}, and Tops{\o}e \cite{Topsoe79}.
The broader subject areas using information projections as a major
component are known as {\em Information Theory and Statistics} and {\em Information
Geometry} (see, e.g., \cite[Chapter~11]{Cover_Thomas}, \cite{CsiszarS_FnT} and references therein).

Given a probability measure $Q$, and a set of probability measures $\mathcal{P}$
defined on an alphabet $\mathcal{A}$, a {\em forward projection} of $Q$ on $\mathcal{P}$
is a $P^*\in\mathcal{P}$ which minimizes $D(P\|Q)$ subject to $P\in\mathcal{P}$.
Forward projections appear predominantly in large deviations theory. By Sanov's
theorem, the exponential decay rate of the probability of rare events is strongly
related to forward projections (see \cite[Theorem~11.4.1]{Cover_Thomas}); furthermore,
in view of the conditional limit theorem, the forward projection of $Q$ on $\mathcal{P}$
arises as the limiting conditional probability measure of a random variable with
distribution $Q \notin \mathcal{P}$, given that the type of its i.i.d. samples
belongs to $\mathcal{P}$ (see \cite[Theorem~11.6.2]{Cover_Thomas}).
The forward projection of a generalization of the relative entropy has been
proposed by Sundaresan in \cite{Sundaresan_ISIT02} and \cite{Sundaresan07}
in the context of guessing under source uncertainty, and it was further studied
in \cite{KumarS15a}.

The R\'{e}nyi divergence, introduced in \cite{Renyientropy} and further studied, e.g.,
in \cite{ErvenH14} and \cite{Shayevitz_ISIT11}, has been investigated so far in various
information-theoretic contexts.
These include generalized cutoff rates and error exponents for hypothesis testing
(e.g., \cite{Csiszar95}), guessing moments (e.g., \cite{ErvenH_ISIT10}), source and
channel coding error exponents (e.g., \cite{Gallager_book1968, Sason15, SGB}),
and other information-theoretic problems.

A motivation for the study of forward projections for the
R\'enyi divergence on some generalized convex sets stems from the following
maximum entropy problem which was proposed by Tsallis in statistical physics
\cite{Tsallis88}, \cite{TsallisMP98}:

 \begin{align}
& \arg\max_{(p_i)} \; S_{\alpha}(P) := \frac{1}{\alpha -1}\Big(1 -
\sum_{i = 1}^W p_i^{\alpha}\Big) \label{tsallis-entropy}\\
& \mbox{subject to } \quad \frac{\sum\limits_{i = 1}^W
p_i^{\alpha}\epsilon_i}{\sum\limits_{i = 1}^W p_i^{\alpha}}
= U^{(\alpha)}, \label{eqn:alpha_linear_constraints}
\end{align}
where $\alpha \in (0,1) \cup (1, \infty)$ is a free parameter, $W$ is the number
of microscopic states, $\{\epsilon_i\}$ are the eigenvalues of the Hamiltonian,
and $U^{(\alpha)}$ is the total internal energy of the system.
The functional $S_{\alpha}(P)$ in \eqref{tsallis-entropy} is known as the {\em Tsallis
entropy}. The constraint in \eqref{eqn:alpha_linear_constraints} is on the {\em escort
probability measure}
\begin{align}
\begin{split}
\label{eq: escort}
& P^{(\alpha)} :=  \bigl(P_1^{(\alpha)}, \ldots, P_W^{(\alpha)} \bigr), \\
& P_i^{(\alpha)} := \frac{p_i^{\alpha}}{\sum\limits_{j=1}^W p_j^{\alpha}}, \quad i\in \{1,\dots,W\}
\end{split}
\end{align}
in contrast to the usual constraint in the Boltzmann-Gibbs statistical physics
\begin{align}
\label{eqn:linear_constraints}
\sum\limits_{i = 1}^W p_i\epsilon_i = U^{(1)}.
\end{align}
The constraint in \eqref{eqn:alpha_linear_constraints}
corresponds to an $\alpha$-linear family (to be formally defined in Section~\ref{sec:PythAndIterate}),
whereas \eqref{eqn:linear_constraints} corresponds to a linear family \cite[Definition 4]{KumarS15b}.
If $Q=U$ is the equiprobable measure on the state space $\{1,\dots,W\}$, then the R\'enyi
divergence $D_{\alpha}(P\|U)$ is related to the objective function $S_{\alpha}(P)$ in \eqref{tsallis-entropy}
via the equation
\begin{align}
D_{\alpha}(P\|U) = \log W + \frac{1}{\alpha-1} \, \log \bigl(1-(\alpha-1) S_{\alpha}(P) \bigr)
\end{align}
which implies that the maximization of $S_{\alpha}(P)$ over the set which is defined in
\eqref{eqn:alpha_linear_constraints} is equivalent to the minimization of $D_{\alpha}(P\|U)$
on the same set of probability measures in \eqref{eqn:alpha_linear_constraints} which
corresponds to an $\alpha$-convex set.

The other problem of interest in this paper is the {\em reverse projection} where the
minimization is over the second argument of the divergence measure.
This problem is intimately related to maximum-likelihood estimation and robust
statistics. Suppose $X_1,\dots,X_n$ are i.i.d. samples drawn according to a probability
measure which is modelled by a parametric family of probability measures
$\Pi = \{P_{\theta} \colon \theta\in\Theta\}$ where $\Theta$ is a parameter space, and
all the members of $\Pi$ are assumed to have a common finite support $\mathcal{A}$.
The maximum-likelihood estimator of the given samples (if it exists) is the minimizer
of $D(\hat P\|P_{\theta})$ subject to $P_{\theta}\in\Pi$, where $\hat P$ is the empirical
probability measure of the observed samples (see, e.g., \cite[Lemma~3.1]{CsiszarS_FnT}).
The minimizing probability measure (if it exists) is called the reverse projection of
$\hat P$ on $\Pi$. Other divergences that have natural connection to statistical
estimation problems include the Hellinger divergence of order $\tfrac{1}{2}$ (see, e.g.,
\cite{Beran77}), Pearson's $\chi^2$-divergence \cite{Pearson1900x}, and so on. All
of these information measures are $f$-divergences (\cite{AliS}, \cite{Csiszar63}) in
the family of Hellinger divergences of order $\alpha \in (0, \infty)$ (note that, up to a
positive scaling factor, Hellinger divergences are equal to the power divergences introduced by
Cressie and Read \cite{CressieR84}). The Hellinger divergences possess a very good
robustness property when a significant fraction of the observed samples are outliers;
the textbooks by Basu et al. \cite{BasuSP} and Pardo \cite{Pardo} address the developments
of studies on inference based on $f$-divergences. Since the R\'enyi divergence
is a monotonically increasing function of the Hellinger divergence (as it
follows from \eqref{eqn:relation-renyi-heli}), minimizing the Hellinger divergence
of order $\alpha \in (0, \infty)$ is equivalent to minimizing the R\'enyi divergence
of the same order. This motivates the study of reverse projections of the R\'{e}nyi
divergence in the context of robust statistics. In \cite[Section~4]{Minka05},
an iterative message-passing algorithm (a.k.a. belief propagation) was used to approximate
reverse projections for the R\'enyi divergence.

In the following, we further motivate our study of forward and reverse
projections for the R\'{e}nyi divergence of order $\alpha \in (0, \infty)$
on $\alpha$-convex sets (note that these terms are formally defined in
Section~\ref{sec:preliminaries}):
\begin{enumerate}[a)]
\item
In view of existing projection theorems for the relative entropy (e.g., \cite{Chentsov68, Csiszar75,
Csiszar84, CsiszarM03}) and Sundaresan's relative $\alpha$-entropy on convex sets
\cite{KumarS15a, KumarS15b}), we study forward and reverse projections
for the R\'enyi divergence of order $\alpha \in (0, \infty)$ on $\alpha$-convex sets.
Our problem reduces to the study of information projections for the relative entropy
on convex sets when $\alpha=1$. Note also that the R\'enyi divergence $D_{\alpha}(P \| Q)$ and
Sundaresan's relative $\alpha$-entropy $\mathscr{I}_{\alpha}(P,Q)$ are related according to the
equality (see \cite[Lemma~2c)]{KumarS15a})
\begin{align}
\label{KS-relation}
\mathscr{I}_{\alpha}(P,Q) = D_{\frac{1}{\alpha}}(P^{(\alpha)} \| Q^{(\alpha)})
\end{align}
where $P^{(\alpha)}$ and $Q^{(\alpha)}$ are, respectively, the associated escort probability
measures of $P$ and $Q$ in \eqref{eq: escort}.

\item
In a recent work \cite{ErvenH14}, van Erven and Harremo\"es proved a Pythagorean
inequality for R\'{e}nyi divergences of order $\alpha \in (0, \infty)$ on $\alpha$-convex
sets under the assumption that the forward projection exists.\footnote{It should be noted
that the R\'{e}nyi divergence does not necessarily satisfy a Pythagorean inequality
on convex sets. For a counter example, see \cite[Appendix~A on p.~19]{ErvenH12}.}
Continuing this study, one of the main objectives of this work is to provide a sufficient
condition for the existence of such a forward projection on an $\alpha$-convex set of
probability measures defined on a general alphabet (see
Theorem~\ref{theorem: existence of forward Dalpha projection}).
Our proof is inspired by the proof of the existence of the forward projection for the
relative $\alpha$-entropy on a convex set (see \cite[Proposition~6]{KumarS15a} and
\cite[Theorem~8]{KumarS15a}).

\item
Forward projections of the relative entropy on linear families and their
orthogonality relationship to exponential families were studied in \cite{Csiszar75}
and \cite[Chapter~3]{CsiszarS_FnT}. We generalize these results by studying
forward projection theorems for the R\'enyi divergence on $\alpha$-linear families.
The solution of this problem yields a parametric family of probability measures which
turns out to be an extension of the exponential family, and it is termed an
{\em $\alpha$-exponential family}. An orthogonality relationship between the
$\alpha$-exponential and $\alpha$-linear families is also established.

\item The orthogonality property of linear and exponential families was used to
transform a reverse projection of relative entropy on an exponential family into
a forward projection on a linear family \cite[Theorem~3.3]{CsiszarS_FnT}.
In this work, we make use of the generalized orthogonality relationship in Item~c)
to transform a reverse projection for the R\'enyi divergence of order $\alpha$ on
an $\alpha$-exponential family into a forward projection on an $\alpha$-linear family.

\item
In \cite[Theorem~3.2]{Csiszar75}, Csisz\'ar proposed a convergent iterative process for
finding the forward projection for the relative entropy on a finite intersection of linear
families. This result is generalized in this work for the R\'enyi divergence of order
$\alpha \in (0, \infty)$ on a finite intersection of $\alpha$-linear families.
\end{enumerate}

The following is an outline of the paper. Section~\ref{sec:preliminaries} provides preliminary
material which is essential to this paper. In Section~\ref{sec:forward-projection}, we study
a sufficient condition for the existence of the forward projection for the R\'enyi divergence on
generalized convex sets. In Section~\ref{sec:PythAndIterate}, we revisit the Pythagorean property
for R\'{e}nyi divergence and prove the iterated projections property as a consequence.
In Section~\ref{sec:ForProjAlphaLinFam}, we establish the form of forward $D_{\alpha}$-projection
on an $\alpha$-linear family and identify the $\alpha$-exponential family as an extension of the
exponential family. In Section~\ref{sec:Orthogonality}, we establish an orthogonality relationship
between the $\alpha$-linear and $\alpha$-exponential families, and in Section~\ref{sec:RevProjAlphaExpFam}
we use this orthogonality property to convert the reverse projection on an $\alpha$-exponential
family into a forward projection on an $\alpha$-linear family. Finally, Section~\ref{sec:summary}
briefly summarizes this paper and provides some concluding remarks.

\section{Preliminaries}
\label{sec:preliminaries}
In this section, we set the notation and formally define terms which
are used in this paper.

Let $(\mathcal{A},\mathscr{F})$ be a measurable space, and let $\mathcal{M}$ denote
the space of all probability measures defined on $\mathcal{A}$.

\begin{definition}
For $P,Q\in \mathcal{M}$, the {\em total variation distance}
between $P$ and $Q$ is defined as
\begin{align}
|P-Q| := 2\sup_{\mathcal{F}\in\mathscr{F}}|P(\mathcal{F}) - Q(\mathcal{F})|.
\end{align}
\end{definition}

\par
If $P$ and $Q$ are absolutely continuous with respect to a common $\sigma$-finite
measure $\mu$ (denoted by $P,Q\ll \mu$), let $p:=\frac{\mathrm{d}P}{\mathrm{d}\mu},
q:=\frac{\mathrm{d}Q}{\mathrm{d}\mu}$ denote their respective densities (Radon-Nikodym
derivatives) with respect to $\mu$ (called $\mu$-densities). Then,
\begin{align}
|P-Q| := \|p - q\|_1 = \int |p - q| \, \mathrm{d}\mu,
\end{align}
and $\mathcal{M}$ together with the total variation distance forms a metric
space. Throughout the paper, the Lebesgue integrals are over the set $\mathcal{A}$.
\vspace{0.1cm}

\par
Pinsker's inequality \cite{Pinsker60} states that
\begin{align}  \label{eq: Pinsker}
\tfrac12 \, |P-Q|^2 \log e \leq D(P \| Q).
\end{align}
Eq.~\eqref{eq: Pinsker} was proved by Csisz\'{a}r \cite{Csiszar67a}
and Kullback \cite{kullbackTV67}, with Kemperman \cite{kemperman} independently
a bit later. From Pinsker's inequality \eqref{eq: Pinsker}, it follows that convergence
in relative entropy also yields convergence in total variation distance (i.e., if
$D(P_n \| P) \to 0$ as $n \to \infty$, then $|P_n - P| \to 0$).

\begin{definition}[R\'enyi divergence]
Let $\alpha \in (0,1)\cup (1,\infty)$. The {\em R\'enyi divergence}
\cite{Renyientropy} of order $\alpha$ from $P$ to $Q$ is given by
\begin{align} \label{RD}
D_{\alpha}(P\|Q) := \frac{1}{\alpha-1} \; \log \left( \int p^{\alpha}
q^{1-\alpha}~\mathrm{d}\mu \right).
\end{align}
If $\alpha =1$, then
\begin{align}
\label{eq: d1}
D_1(P\|Q) := D(P\|Q),
\end{align}
which is the continuous extension of $D_{\alpha}(P \| Q)$ at $\alpha=1$.
\end{definition}

\begin{definition} \label{def: Hellinger}
The Hellinger divergence \cite[Definition~2.10]{LieseV_book87} of order
$\alpha \in (0,1) \cup (1, \infty)$ from $P$ to $Q$ is given by
\begin{align}
\label{eq: Hel-divergence}
\mathscr{H}_{\alpha}(P \| Q) := \frac{1}{\alpha-1} \, \left(\int p^{\alpha}
q^{1-\alpha}~\mathrm{d}\mu - 1\right).
\end{align}
The continuous extension of $\mathscr{H}_{\alpha}(P \| Q)$ at $\alpha=1$ yields
\begin{align} \label{eq: Hel-divergence- order1}
\mathscr{H}_1(P \| Q) \, \log e = D(P \| Q).
\end{align}
\end{definition}

Note that $|P-Q|$, $D_{\alpha}(P \| Q)$ and $\mathscr{H}_{\alpha}(P\|Q)$ are
non-negative, and are equal to zero if and only if $P=Q$. These measures can
be expressed in terms of $f$-divergences \cite{AliS, Csiszar63, Csiszar67a},
and they do not depend on the choice of the reference measure $\mu$. Note
that, from \eqref{RD} and \eqref{eq: Hel-divergence},
\begin{eqnarray}
\label{eqn:relation-renyi-heli}
D_{\alpha}(P\|Q) = \frac{1}{\alpha-1} \, \log \bigl(1 + (\alpha -1)\mathscr{H}_{\alpha}(P\|Q)\bigr),
\end{eqnarray}
showing that the R\'enyi divergence is monotonically increasing with the Hellinger divergence.

\begin{definition}[$(\alpha,\lambda)$-mixture \cite{ErvenH14}]
Let $P_0,P_1\ll \mu$, let $\alpha\in (0,\infty)$, and let $\lambda \in (0,1)$. The
{\em $(\alpha,\lambda)$-mixture} of $(P_0, P_1)$ is the probability
measure $S_{0,1}$ with $\mu$-density
\begin{align}
\label{eqn:alpha-lambda-mixture}
s_{0,1} := \frac{1}{Z} \Bigl[(1-\lambda)p_0^{\alpha}+\lambda p_1^{\alpha}\Bigr]^{\frac1{\alpha}},
\end{align}
where $Z$ is a normalizing constant such that $\int s_{0,1}~d\mu = 1$, i.e.,
\begin{align} \label{eq: z}
Z = \int \Bigl[(1-\lambda)p_0^{\alpha}+\lambda p_1^{\alpha}\Bigr]^{\frac1{\alpha}} \, \mathrm{d}\mu.
\end{align}
\end{definition}
Here, for simplicity, we suppress the dependence of $S_{0,1}$ and $Z$ on $\alpha,\lambda$.
Note that $s_{0,1}$ is well-defined as $Z$ is always positive and finite. Indeed,
for $\alpha\in (0,\infty)$ and $\lambda \in [0,1]$,
\begin{align}
0 \leq \Bigl[(1-\lambda)p_0^{\alpha}+\lambda p_1^{\alpha}\Bigr]^{\frac1{\alpha}}
\leq \max\{p_0,p_1\}\le p_0+p_1
\end{align}
which implies that $0 < Z \leq 2$. From \eqref{eqn:alpha-lambda-mixture}, for
$\lambda \in [0,1]$, the $(\alpha,\lambda)$-mixture of $(P_0,P_1)$
is the same as the $(\alpha,1-\lambda)$-mixture of $(P_1,P_0)$.

\begin{definition}[$\alpha$-convex set]
Let $\alpha \in (0,\infty)$. A set of probability measures $\mathcal{P}$
is said to be {\em $\alpha$-convex} if, for every $P_0, P_1\in \mathcal{P}$ and
$\lambda \in (0,1)$, the $(\alpha,\lambda)$-mixture $S_{0,1}\in \mathcal{P}$.
\end{definition}

\section{Existence of Forward $D_{\alpha}$-Projections on $\alpha$-Convex Sets}
\label{sec:forward-projection}
In this section, we define what we mean by a forward $D_{\alpha}$-projection, and
then provide a sufficient condition for the existence of forward $D_{\alpha}$-projections
on $\alpha$-convex sets.
\vspace{0.1cm}

\begin{definition}[Forward $D_{\alpha}$-projection]
Let $Q\in \mathcal{M}$, $\mathcal{P}\subseteq\mathcal{M}$, and $\alpha \in (0, \infty)$.
If there exists $P^*\in \mathcal{P}$ which attains the global minimum of $D_{\alpha}(P\|Q)$
over all $P\in \mathcal{P}$ and $D_{\alpha}(P^*\|Q) < \infty$, then $P^*$ is said to be a
{\em forward $D_{\alpha}$-projection} of $Q$ on $\mathcal{P}$.
\end{definition}
\vspace{0.1cm}

We next proceed to show the existence of a forward $D_{\alpha}$-projection on an $\alpha$-convex
set. It has been shown in \cite[Theorem~14]{ErvenH14} that if $\mathcal{P}$ is an
$\alpha$-convex set and $P^*$ exists, then the {\em Pythagorean inequality} holds, i.e.,
\begin{eqnarray}
\label{pythagorean-inequality}
D_{\alpha}(P\|Q)\ge D_{\alpha}(P\|P^*)+D_{\alpha}(P^*\|Q), \quad \forall \, P \in \mathcal{P}.
\end{eqnarray}
However, the existence of the forward $D_{\alpha}$-projection was not addressed in \cite{ErvenH14}.
We show that if the $\alpha$-convex set
$\mathcal{P}$ is closed with respect to the total variation distance, then the forward
$D_{\alpha}$-projection exists. The proof is inspired by the proof of the existence of a
forward projection for the relative $\alpha$-entropy on a convex set \cite[Theorem~8]{KumarS15a}.
Before getting to the main result of this section, we prove the following inequality for the
Hellinger divergence.
\vspace{0.05cm}

\begin{lemma}[Apollonius theorem for the Hellinger divergence]
\label{thm:apollonius-theorem}
If $\alpha \in (1,\infty)$, $\lambda \in (0,1)$, and $P_0,P_1,Q$ are probability measures where
$P_0,P_1,Q\ll \mu$, then
\begin{align}
\label{eqn:apollonius-theorem}
& (1-\lambda) \bigl(\mathscr{H}_{\alpha}(P_0\|Q) -
\mathscr{H}_{\alpha}(P_0\|S_{0,1}) \bigr) \nonumber \\
& + \lambda \bigl(\mathscr{H}_{\alpha}(P_1\|Q)
- \mathscr{H}_{\alpha}(P_1\|S_{0,1}) \bigr)
\ge \mathscr{H}_{\alpha}(S_{0,1}\|Q),
\end{align}
and the inequality in \eqref{eqn:apollonius-theorem} is reversed for $\alpha\in (0,1)$.
\end{lemma}

\begin{IEEEproof}
The left side of \eqref{eqn:apollonius-theorem} simplifies to
\begin{align}
& \bigl(1-\lambda)(\mathscr{H}_{\alpha}(P_0\|Q) -
\mathscr{H}_{\alpha}(P_0\|S_{0,1}) \bigr)
+ \lambda \bigl(\mathscr{H}_{\alpha}(P_1\|Q)
- \mathscr{H}_{\alpha}(P_1\|S_{0,1}) \bigr)\nonumber\\[0.1cm]
& = \frac{1-\lambda}{\alpha-1} \, \int p_0^{\alpha}
\bigl(q^{1-\alpha}-s_{0,1}^{1-\alpha} \bigr)~\mathrm{d}\mu  +
\frac{\lambda}{\alpha-1} \, \int p_1^{\alpha} \bigl(q^{1-\alpha}
-s_{0,1}^{1-\alpha}\bigr)~\mathrm{d}\mu \nonumber\\[0.1cm]
& = \frac{1}{\alpha-1}\int \bigl((1-\lambda)p_0^{\alpha}
+ \lambda p_1^{\alpha} \bigr)
\bigl(q^{1-\alpha}-s_{0,1}^{1-\alpha}\bigr)~\mathrm{d}\mu\nonumber\\[0.1cm]
& = \frac{1}{\alpha-1}\int Z^{\alpha} \, s_{0,1}^{\alpha}
\, \bigl(q^{1-\alpha}-s_{0,1}^{1-\alpha}\bigr)~\mathrm{d}\mu\nonumber\\[0.1cm]
& =  \frac{Z^\alpha}{\alpha-1}  \left( \int s_{0,1}^\alpha \, q^{1-\alpha}
\, \mathrm{d}\mu - 1 \right) \nonumber\\[0.1cm]
& = Z^{\alpha} \;  \mathscr{H}_{\alpha}(S_{0,1}\|Q).  \label{eq: new identity}
\end{align}
The result follows since, by invoking Jensen's inequality to \eqref{eq: z} (see
\cite[Lemma 3]{ErvenH14}), $Z\ge 1$ if $\alpha\in (1,\infty)$, and $0<Z\le 1$
if $\alpha\in (0,1)$.
\end{IEEEproof}
\vspace{0.1cm}

\begin{remark}
Lemma~\ref{thm:apollonius-theorem} is analogous to the Apollonius theorem for the relative
$\alpha$-entropy \cite[Proposition~6]{KumarS15a} where $S_{0,1}$ is replaced by a convex
combination of $P_0$ and $P_1$. In view of \eqref{eq: Hel-divergence- order1}
and since $Z=1$ when $\alpha=1$ (see \eqref{eq: z}), it follows that
\eqref{eq: new identity} reduces to the {\em parallelogram law} for the relative
entropy \cite[(2.2)]{Csiszar75} when $\alpha =1$ and $\lambda = \tfrac12$.
\end{remark}

We are now ready to state our first main result.
\begin{theorem}[Existence of forward $D_{\alpha}$-projection]
\label{theorem: existence of forward Dalpha projection}
Let $\alpha\in (0,\infty)$, and let $Q$ be an arbitrary probability measure
defined on a set $\mathcal{A}$. Let $\mathcal{P}$ be an $\alpha$-convex set
of probability measures defined on $\mathcal{A}$, and assume that $\mathcal{P}$
is closed with respect to the total variation distance. If there exists
$P\in \mathcal{P}$ such that $D_{\alpha}(P\|Q)<\infty$, then there exists
a forward $D_{\alpha}$-projection of $Q$ on $\mathcal{P}$.
\end{theorem}
\vspace{0.1cm}

\begin{IEEEproof}
We first consider the case where $\alpha\in (1,\infty)$. Let $\{P_n\}$
be a sequence in $\mathcal{P}$ such that $D_{\alpha}(P_n\|Q)<\infty$ and
$D_{\alpha}(P_n\|Q)\to \inf_{P\in \mathcal{P}}D_{\alpha}(P\|Q) =:
D_{\alpha}(\mathcal{P}\|Q)$. Then, in view of  \eqref{eqn:relation-renyi-heli},
$\mathscr{H}_{\alpha}(P_n\|Q)<\infty$ and $\mathscr{H}_{\alpha}(P_n\|Q) \to
\inf_{P\in \mathbb{E}}\mathscr{H}_{\alpha}(P\|Q) =: \mathscr{H}_{\alpha}(\mathcal{P}\|Q)$.

Let $m,n \in \mathbb{N}$, and let $S_{m,n}$ be the $(\alpha,\lambda)$-mixture of $(P_m, P_n)$,
i.e., $S_{m,n}$ is the probability measure with $\mu$-density
\begin{align}
s_{m,n} = \frac{1}{Z_{m,n}} \, \Bigl[(1-\lambda)p_m^{\alpha} + \lambda p_n^{\alpha}\Bigr]^{1/\alpha},
\end{align}
where $Z_{m,n}$ is the normalizing constant such that $\int s_{m,n}~d\mu = 1$.
Applying Lemma~\ref{thm:apollonius-theorem}, we have
\begin{align}
\label{eq1: limit-infimum-non-negative}
0 \le & (1-\lambda) \mathscr{H}_{\alpha}(P_m\|S_{m,n}) + \lambda
\mathscr{H}_{\alpha}(P_n\|S_{m,n})\\
\label{eq2: limit-infimum-non-negative}
\le & (1-\lambda) \mathscr{H}_{\alpha}(P_m\|Q) + \lambda
\mathscr{H}_{\alpha}(P_n\|Q) -  \mathscr{H}_{\alpha}(S_{m,n}\|Q).
\end{align}
Since $\mathscr{H}_{\alpha}(P_n\|Q)\to \mathscr{H}_{\alpha}(\mathcal{P}\|Q)$ as we let
$n \to \infty$, and $\mathscr{H}_{\alpha}(S_{m,n}\|Q)\ge \mathscr{H}_{\alpha}(\mathcal{P}\|Q)$
(note that $S_{m,n}\in \mathcal{P}$ due to the $\alpha$-convexity of the set
$\mathcal{P}$), the limit supremum of the right side of \eqref{eq2: limit-infimum-non-negative}
is non-positive as $n,m \to \infty$. From the left side of \eqref{eq1: limit-infimum-non-negative},
the limit infimum of the right side of \eqref{eq2: limit-infimum-non-negative} is also non-negative.
This implies that the limit of the right side of \eqref{eq2: limit-infimum-non-negative} is zero,
which also implies that the right side of \eqref{eq1: limit-infimum-non-negative} converges to zero
as we let $m,n \to \infty$;
consequently, $\mathscr{H}_{\alpha}(P_n\|S_{m,n})\to 0$ and $\mathscr{H}_{\alpha}(P_m\|S_{m,n})
\to 0$ as $m,n\to\infty$. Since the Hellinger divergence, $\mathscr{H}_{\alpha}(\cdot \| \cdot)$,
is monotonically increasing in $\alpha$
\cite[Proposition~2.7]{LieseV_book87}\footnote{A simple proof of the monotonicity of the
Hellinger divergence in $\alpha$ appears in \cite[Theorem~33]{SasonV15}.}, it follows
from \eqref{eq: Hel-divergence- order1} that $D(P_n\|S_{m,n})\to 0$ and
$D(P_m\|S_{m,n})\to 0$ as $m,n\to\infty$, which, in turn implies (via Pinsker's
inequality \eqref{eq: Pinsker}) that $|P_n - S_{m,n}|\to 0$ and $|P_m - S_{m,n}|\to 0$ as $m, n\to \infty$.
The triangle inequality for the total variation distance yields that
$|P_n-P_m| \to 0$ as $m, n\to \infty$, i.e., $\{P_n\}$ is a Cauchy sequence
in $\mathcal{P}$, which therefore converges to some $P^*\in \mathcal{P}$ due to the completeness
of $\mathcal{P}$ with respect to the total variation distance. Subsequently,
the corresponding sequence of $\mu$-densities $\{p_n\}$ converges to the $\mu$-density
$p^*$ in $L^1$; this implies that there exists a sub-sequence $\{p_{n_k}\}$
which converges $\mu$-almost everywhere (a.e.) to $p^*$. By Fatou's lemma and
\eqref{eq: Hel-divergence}, it follows that for $\alpha \in (1, \infty)$
\begin{align}
\mathscr{H}_{\alpha}(\mathcal{P}\|Q) &= \lim_{n\to\infty}\mathscr{H}_{\alpha}(P_n\|Q) \nonumber \\
& = \lim_{k\to\infty}\mathscr{H}_{\alpha}(P_{n_k}\|Q) \nonumber \\
& \geq \mathscr{H}_{\alpha}(P^*\|Q)
\end{align}
which implies that $P^*$ is a forward $\mathscr{H}_{\alpha}$-projection of $Q$ on $\mathcal{P}$.
In view of \eqref{eqn:relation-renyi-heli}, this is equivalent to saying
that $P^*$ is a forward $D_{\alpha}$-projection of $Q$ on $\mathcal{P}$.

We next consider the case where $\alpha\in (0,1)$. Abusing notation a little, we use
the same letter $\mathcal{P}$ to denote a set of probability measures as well as the
set of their corresponding $\mu$-densities. Since $\alpha <1$,
\begin{align}
\label{supremum-over-P}
\inf_{P\in \mathcal{P}}D_{\alpha}(P\|Q) & = \frac{1}{\alpha-1}\log
\left(\sup_{p\in\mathcal{P}}\int p^{\alpha}q^{1-\alpha}~\mathrm{d}\mu \right)\\
\label{supremum-over-P-hat}
& = \frac{1}{\alpha-1}\log\left(\sup_{g\in\widehat{\mathcal{P}}}\int
gh~\mathrm{d}\mu\right),
\end{align}
where $g := s p^{\alpha}, h:=q^{1-\alpha}$ and
\begin{align} \label{eq: set of hat P}
\widehat{\mathcal{P}}:=\{sp^{\alpha}\colon p\in \mathcal{P}, ~0\le s\le 1\}.
\end{align}
Notice that the multiplication of $p^{\alpha}$ by the scalar
$s\in [0,1]$ in the right side of \eqref{eq: set of hat P}
does not affect the supremum in \eqref{supremum-over-P-hat}.
This supremum, if attained, is obtained by some
$g = sp^{\alpha}$ with $s=1$ and $p \in \mathcal{P}$.
The purpose of introducing $s\in [0,1]$ is to make the
optimization in \eqref{supremum-over-P-hat} over a convex
set (as it is shown in the sequel).

Let $\beta = \frac{1}{\alpha}$ and $\beta' := \frac{1}{1-\alpha}$; note
that $\beta$ and $\beta'$ are H\"{o}lder
conjugates (i.e., $\tfrac1{\beta} + \tfrac1{\beta'} = 1$).
Then, $\int h^{\beta'}~\mathrm{d}\mu = \int q~\mathrm{d}\mu = 1$, so
$h\in L^{\beta'}(\mu)$. By invoking H\"{o}lder's inequality,
it follows that $F_h(g) := \int gh~\mathrm{d}\mu$ is a continuous linear functional on
$L^{\beta}(\mu)$. Thus, the supremum is of a continuous linear functional on the reflexive
Banach space $L^{\beta}(\mu)$. We claim that $\widehat{\mathcal{P}}$ is closed and convex
in $L^{\beta}(\mu)$. For the moment, we assume that the claim holds, and later prove it.
A convex set which is closed with respect to the norm topology is also closed with respect
to the weak topology \cite[Ch.~10,~Cor.~23]{Royden_book88}. Note that the weak topology
on $L^{\beta}(\mu)$ is the smallest topology on $L^{\beta}(\mu)$ for which the continuity
of the linear functionals on $L^{\beta}(\mu)$ is preserved. Moreover, for any
$g=sp^{\alpha}\in \widehat{\mathcal{P}}$, $\|g\|_{\beta} = s\le 1$. Hence, $\widehat{\mathcal{P}}$
is a subset of the unit sphere of $L^{\beta}(\mu)$.
By the Banach-Alaoglu theorem \cite[Ch.~10,~Th.~17]{Royden_book88} and the fact that
$L^{\beta}(\mu)$ is a reflexive Banach space, it follows that the unit sphere
$\{g\colon \|g\|_{\beta}\le 1\}$ is compact with respect to the weak topology of $L^{\beta}$.
Hence, $\widehat{\mathcal{P}}$ is a closed subset of a compact set with respect to the weak
topology of $L^{\beta}(\mu)$, so $\widehat{\mathcal{P}}$ is also compact in the weak topology.
Thus, the supremum in \eqref{supremum-over-P-hat} is of a continuous linear functional over
a compact set in $L^{\beta}(\mu)$, which yields that this supremum is attained.

To complete the proof for $\alpha \in (0,1)$, we prove the claim that
$\widehat{\mathcal{P}}$ is convex and closed.
To verify that $\widehat{\mathcal{P}}$ is convex, let
$s_1p_1^{\alpha}, s_2p_2^{\alpha}\in \widehat{\mathcal{P}}$ and
$\lambda\in (0,1)$. We can write
$\lambda s_1p_1^{\alpha}+(1-\lambda)s_2p_2^{\alpha} = sp^{\alpha}$ with
\begin{align}
p = \frac{1}{Z}\left(\frac{\lambda s_1p_1^{\alpha}
+(1-\lambda)s_2p_2^{\alpha}}{\lambda s_1 + (1-\lambda)s_2}\right)^{1/\alpha},
\end{align}
where $Z$ is the normalizing constant, and
$s = \bigl( \lambda s_1 +(1-\lambda)s_2 \bigr) Z^{\alpha}$.
For $\alpha\in (0,1)$, $0 < Z \le 1$ by \cite[Lemma 3]{ErvenH14}
which implies that $s \in [0,1]$. This proves the convexity
of $\widehat{\mathcal{P}}$.

Next, to prove that $\widehat{\mathcal{P}}$ is closed, let
$g_n:=s_n p_n^{\alpha}\in \widehat{\mathcal{P}}$ be such that
$g_n\to g$ in $L^{\beta}(\mu)$. We need to show that
$g\in \widehat{\mathcal{P}}$. Since
$s_n = \|g_n\|_{\beta} \to \|g\|_{\beta}$, we have
$\|g\|_{\beta}\le 1$. If $\|g\|_{\beta}=0$, then $g=0$
$\mu$-a.e., and hence obviously
$g\in\widehat{\mathcal{P}}$. Since $\beta = \tfrac1{\alpha} > 1$,
it follows that if $\|g\|_{\beta}>0$, then
$p_n^{\alpha}=g_n/\|g_n\|_{\beta}\to g/\|g\|_{\beta}$
in $L^{\beta}(\mu)$, and therefore
$p_n\to (g/\|g\|_{\beta})^{\beta}$ in $L^1(\mu)$.\footnote{If
$\beta>1$ and $\{f_n\}$ converges to $f$ in $L^{\beta}$, then
an application of the mean-value theorem and H\"{o}lder's
inequality yields $\bigl\| |f_n|^{\beta} - |f|^{\beta} \bigr\|
\leq \beta ( \| f_n \|_{\beta} + \|f\|_{\beta})^{\beta-1} \, \|f_n - f\|_1$;
hence, $\{|f_n|^{\beta}\}$ converges to $|f|^\beta$ in $L^1$.
Since non-negative functions are considered in our case, we can ignore the
absolute values so $\{f_n^{\beta}\}$ converges to $f^\beta$ in $L^1$.}
Since $\mathcal{P}$ is closed in $L^1(\mu)$, we have
$g/\|g\|_{\beta} = p^* \in \mathcal{P}$, and
$g = \|g\|_{\beta} \cdot p^*\in \widehat{\mathcal{P}}$.

\end{IEEEproof}

\begin{remark}
The fact underlying the above proof is that the maximum or minimum of a
continuous function over a compact set is always attained. Although the
actual set $\mathcal{P}$ in \eqref{supremum-over-P}, over which we wish
to optimize the functional, is not compact, it was possible to modify it
into the set $\widehat{\mathcal{P}}$ in \eqref{eq: set of hat P} without
affecting the optimal value in \eqref{supremum-over-P-hat}; the modified
set $\widehat{\mathcal{P}}$ was compact in an appropriate topology
where the functional also remains to be continuous.
\end{remark}

\newpage
\section{The Pythagorean Property and Iterated Projections}
\label{sec:PythAndIterate}
In this section we first revisit the Pythagorean property for a finite
alphabet and use it to prove a convergence theorem for iterative projections.
Throughout this section, we assume that the probability measures are
defined on a {\em finite set} $\mathcal{A}$. For a probability measure
$P$, let its support be given by $\text{Supp}(P) := \{a\in \mathcal{A}\colon P(a)>0\}$;
for a set of probability measures $\mathcal{P}$, let
\begin{align}
\text{Supp}(\mathcal{P}) := \bigcup_{P\in\mathcal{P}}\text{Supp}(P).
\end{align}
Let us first recall the Pythagorean property for a R\'enyi divergence
on an $\alpha$-convex set. As it is in the cases of relative entropy
\cite{CsiszarS_FnT} and relative $\alpha$-entropy \cite{KumarS15b}, the
Pythagorean property is crucial in establishing orthogonality properties.
In the sequel, we assume that $Q$ is a probability measure with
$\text{Supp}(Q) = \mathcal{A}$.

\begin{proposition}[The Pythagorean property]
\label{prop:pythagorean-property}
Let $\alpha \in (0,1)\cup (1,\infty)$, let $\mathcal{P}\subseteq \mathcal{M}$
be an $\alpha$-convex set, and $Q\in \mathcal{M}$.
\begin{enumerate}[a)]
 \item If $P^*$ is a forward $D_{\alpha}$-projection of $Q$ on $\mathcal{P}$,
 then
\begin{eqnarray}
\label{pythagorean-inequality1}
D_{\alpha}(P\|Q)\ge D_{\alpha}(P\|P^*)+D_{\alpha}(P^*\|Q), \quad
\forall \, P \in \mathcal{P}.
\end{eqnarray}
Furthermore, if $\alpha >1$, then $\text{Supp}(P^*) = \text{Supp}(\mathcal{P})$.
\item Conversely, if \eqref{pythagorean-inequality1} is satisfied for some
$P^*\in\mathcal{P}$, then $P^*$ is a forward $D_{\alpha}$-projection of $Q$
on $\mathcal{P}$.
\end{enumerate}
\end{proposition}
\vspace{0.1cm}

\begin{IEEEproof}
a) In view of the proof of \cite[Theorem~14]{ErvenH14}, for every $P \in \mathcal{P}$
and $t \in [0,1]$, let $P_t \in \mathcal{P}$ be the $(\alpha, t)$-mixture of
$(P^*, P)$; since $D_{\alpha}(P_t \| Q)$ is minimized at $t=0$, then (see \cite[pp.~3806--3807]{ErvenH14}
for detailed calculations)
\begin{align}
0 & \leq \frac{\mathrm{d}}{\mathrm{d}t} \, D_{\alpha}(P_t \| Q) \Big|_{t=0} \nonumber \\[0.1cm]
\label{eqn:derivative-calculation}
& = \frac{1}{\alpha - 1}\left(\frac{\sum_a P(a)^{\alpha}Q(a)^{1-\alpha}}{\sum_a P^*(a)^{\alpha}Q(a)^{1-\alpha}}
- \sum_a P(a)^{\alpha}P^*(a)^{1-\alpha}\right)
\end{align}
which is equivalent to \eqref{pythagorean-inequality1}. To show that
$\text{Supp}(P^*) = \text{Supp}(\mathcal{P})$ for $\alpha > 1$,
suppose that there exist $P\in \mathcal{P}$ and $a\in\mathcal{A}$ such that
$P^*(a) = 0$ but $P(a)>0$. Then \eqref{eqn:derivative-calculation} is violated
since its right side is equal, in this case, to $-\infty$ (recall that by assumption
$\text{Supp}(Q)=\mathcal{A}$ so, if $\alpha > 1$,
$\sum_a P(a)^{\alpha}Q(a)^{1-\alpha}, \; \sum_a P^*(a)^{\alpha}Q(a)^{1-\alpha} \in (0, \infty)$,
and $\sum_a P(a)^{\alpha}P^*(a)^{1-\alpha} = +\infty$). This contradiction proves the last
assertion in Proposition~\ref{prop:pythagorean-property}a).

b) From \eqref{pythagorean-inequality1}, we have
\begin{align}
D_{\alpha}(P\|Q) & \ge D_{\alpha}(P\|P^*) + D_{\alpha}(P^*\|Q) \nonumber \\
\label{eq: Phytagorean}
& \ge  D_{\alpha}(P^*\|Q) \quad \forall P\in \mathcal{P}.
\end{align}
\end{IEEEproof}
\vspace{0.1cm}

\begin{remark}
The Pythagorean property \eqref{pythagorean-inequality1} holds for probability
measures defined on a general alphabet $\mathcal{A}$, as it is proved in
\cite[Theorem~14]{ErvenH14}.
The novelty in Proposition~\ref{prop:pythagorean-property} is in the last assertion
of a), extending the result for the relative entropy in \cite[Theorem~3.1]{CsiszarS_FnT},
for which $\mathcal{A}$ needs to be a finite set.
\end{remark}
\vspace{0.1cm}

\begin{corollary} \label{corollary: uniqueness}
Let $\alpha \in (0,\infty)$. If a forward $D_{\alpha}$-projection on an
$\alpha$-convex set exists, then it is unique.
\end{corollary}
\vspace{0.1cm}

\begin{IEEEproof}
For $\alpha =1$, since an $\alpha$-convex set is particularized to a convex set,
this result is known in view of \cite[p.~23]{CsiszarS_FnT}. Next, consider the
case where $\alpha \in (0,1) \cup (1, \infty)$. Let $P_1^*$ and $P_2^*$ be
forward $D_{\alpha}$-projections of $Q$ on an $\alpha$-convex set $\mathcal{P}$.
Applying Proposition~\ref{prop:pythagorean-property}, we have
\begin{eqnarray*}
D_{\alpha}( P_2^*\|Q)\ge D_{\alpha}( P_2^*\|P_1^*)+D_{\alpha}(P_1^*\|Q).
\end{eqnarray*}
Since $D_{\alpha}(P_1^*\|Q) = D_{\alpha}(P_2^*\|Q)$, we must have
$D_{\alpha}(P_2^*\|P_1^*) = 0$ which yields $P_1^* = P_2^*$.
\end{IEEEproof}
\vspace{0.1cm}

The last assertion in Proposition~\ref{prop:pythagorean-property}a) shows that
$\text{Supp}(P^*) = \text{Supp}(\mathcal{P})$ if $\alpha \in (1, \infty)$. The
following counterexample illustrates that this equality does not necessarily hold
for $\alpha\in (0,1)$.
\begin{example}
\label{eg:counterexample}
Let $\mathcal{A} = \{1,2,3,4\}$, \, $\alpha = \tfrac12$, \,
$f \colon \mathcal{A} \to \mathbb{R}$ be given by
\begin{align}
\label{eq0: example1}
& f(1)=1, \; f(2)=-3, \; f(3)=-5, \; f(4)=-6
\end{align}
and let $Q(a) = \tfrac14$ for all $a \in \mathcal{A}$.
Consider the following $\alpha$-convex set:\footnote{This set is
characterized in \eqref{eqn:linear-family} as an $\alpha$-linear family.}
\begin{align} \label{eq: set P in example}
\mathcal{P} := \biggl\{P\in\mathcal{M} : \sum_{a \in \mathcal{A}} P(a)^{\alpha}f(a) = 0 \biggr\}.
\end{align}
Let
\begin{align}
P^*(1) = \tfrac{9}{10}, \; P^*(2) = \tfrac1{10}, \; P^*(3) = 0, \; P^*(4)=0.
\end{align}
It is easy to check that $P^* \in \mathcal{P}$.
Furthermore, setting $\theta^* = \tfrac15$ and $Z = \tfrac25$ yields
\begin{align}
\label{eq1: example1}
& P^*(a)^{1-\alpha} = Z^{\alpha-1}\Big[Q(a)^{1-\alpha} + (1-\alpha) \, f(a) \,
\theta^* \Big],
\end{align}
for all $ a \in \{1, 2, 3\} $, and
\begin{align}
\label{eq2: example1}
& P^*(4)^{1-\alpha} > Z^{\alpha-1}\Big[Q(4)^{1-\alpha} + (1-\alpha) \, f(4)
\, \theta^*\Big].
\end{align}
From \eqref{eq: set P in example}, \eqref{eq1: example1} and \eqref{eq2: example1}, it
follows that for every $P\in \mathcal{P}$
\begin{align}
\label{eq3: example1}
\sum_{a \in \mathcal{A}} P(a)^{\alpha}P^*(a)^{1-\alpha} \ge Z^{\alpha -1}
\sum_{a \in \mathcal{A}} P(a)^{\alpha}Q(a)^{1-\alpha}.
\end{align}
Furthermore, it can be also verified that
\begin{align}
\label{eq4: example1}
Z^{\alpha -1} \sum_{a \in \mathcal{A}} P^*(a)^{\alpha}Q(a)^{1-\alpha} = 1.
\end{align}
Assembling \eqref{eq3: example1} and \eqref{eq4: example1} yields
\begin{align}
\label{eq5: example1}
\sum_{a \in \mathcal{A}} P(a)^{\alpha}P^*(a)^{1-\alpha} \ge \frac{
\sum_{a \in \mathcal{A}} P(a)^{\alpha}Q(a)^{1-\alpha}}{\sum_{a
\in \mathcal{A}} P^*(a)^{\alpha}Q(a)^{1-\alpha}},
\end{align}
which is equivalent to \eqref{pythagorean-inequality1}. Hence,
Proposition~\ref{prop:pythagorean-property}b) implies that $P^*$
is the forward $D_{\alpha}$-projection of $Q$ on $\mathcal{P}$.
Note, however, that $\text{Supp}(P^*) \neq \text{Supp}(\mathcal{P})$;
to this end, from \eqref{eq: set P in example}, it can be verified
numerically that
\begin{align}
\label{eq6: example1}
P = (0.984688, \, 0.005683, \, 0.004180, \, 0.005449) \in \mathcal{P}
\end{align}
which implies that $\text{Supp}(P^*) = \{1,2\}$ whereas
$\text{Supp}(\mathcal{P}) = \{1, 2, 3, 4\}$.
\end{example}

\begin{definition}[$\alpha$-linear family]
\label{alpha-linear-family}
Let $\alpha \in (0,\infty)$, and $f_1,\ldots,f_k$
be real-valued functions defined on $\mathcal{A}$.
The \emph{$\alpha$-linear family} determined by $f_1, \dots , f_k$
is defined to be the following parametric family of probability
measures defined on $\mathcal{A}$:
\begin{align}
\label{eqn:alpha-linear-family}
\mathscr{L}_{\alpha} := \left\{ P \in \mathcal{M} \colon P(a)
= \left[\sum\limits_{i=1}^k \theta_i f_i(a)\right]^{\frac{1}{\alpha}},
\quad (\theta_1,\dots,\theta_k)\in\mathbb{R}^k \right\}.
\end{align}
\end{definition}

For typographical convenience, we have suppressed the dependence of
$\mathscr{L}_{\alpha}$ in $f_1,\dots, f_k$. It is easy to see that
$\mathscr{L}_{\alpha}$ is an $\alpha$-convex set.
Without loss of generality, we shall assume that $f_1,\dots ,f_k$,
as $|\mathcal{A}|$-dimensional vectors, are mutually orthogonal
(otherwise, by the Gram-Schmidt procedure, these vectors can be
orthogonalized without affecting the corresponding $\alpha$-linear
family in \eqref{eqn:alpha-linear-family}).
Let $\mathcal{F}$ be the subspace of $\mathbb{R}^{|\mathcal{A}|}$
spanned by $f_1,\dots,f_k$, and let $\mathcal{F}^{\perp}$ denote
the orthogonal complement of $\mathcal{F}$. Hence, there exist
$f_{k+1},\dots, f_{|\mathcal{A}|}$ such that $f_1, \ldots,
f_{|\mathcal{A}|}$ are mutually orthogonal as $|\mathcal{A}|$-dimensional
vectors, and $\mathcal{F}^{\perp} = \text{Span}\{f_{k+1}, \ldots,
f_{|\mathcal{A}|}\}$. Consequently, from \eqref{eqn:alpha-linear-family},
\begin{align}
\label{eqn:linear-family}
\mathscr{L}_{\alpha} = \left\{P\in\mathcal{M} \colon
\sum \limits_a P(a)^{\alpha} f_i(a) = 0,
\quad \forall \, i\in \{k+1,\dots, |\mathcal{A}|\} \right\}.
\end{align}
From \eqref{eqn:linear-family}, the set $\mathscr{L}_{\alpha}$ is closed.
We shall now focus our attention on forward $D_{\alpha}$-projections on
$\alpha$-linear families.
\vspace{0.2cm}

\begin{theorem}[Pythagorean equality]
\label{thm:pythagorean-equality}
Let $\alpha >1$, and let $P^*$ be the forward $D_{\alpha}$-projection
of $Q$ on $\mathscr{L}_{\alpha}$. Then, $P^*$ satisfies
\eqref{pythagorean-inequality1} with equality, i.e.,
\begin{eqnarray}
\label{eqn:pythagorean-equality}
D_{\alpha}(P\|Q) = D_{\alpha}(P\|P^*) + D_{\alpha}(P^*\|Q),
\quad \forall \, P \in \mathscr{L}_{\alpha}.
\end{eqnarray}
\end{theorem}
\vspace{0.1cm}

\begin{IEEEproof}
For $t\in [0,1]$ and $P\in\mathscr{L}_{\alpha}$, let $P_t$ be the $(\alpha,t)$-mixture of
$(P, P^*)$, i.e.,
\begin{align} \label{eqn:s-t}
P_t(a) = \frac1{Z_t} \, \Bigl[(1-t){P^*(a)}^{\alpha} + t P(a)^{\alpha}\Bigr]^{\frac{1}{\alpha}},
\end{align}
where
\begin{align} \label{eqn:z-t}
Z_t := \sum\limits_a\left[(1-t){P^*(a)}^{\alpha} + t P(a)^{\alpha}\right]^{\frac{1}{\alpha}}.
\end{align}
Since $P_t\in\mathcal{P}$,
\begin{eqnarray}
\label{eqn:nonnegative}
D_{\alpha}(P_t\|Q) \ge D_{\alpha}(P^*\|Q) = D_{\alpha}(P_0\|Q),
\end{eqnarray}
which yields
\begin{eqnarray}
\label{eqn:derivative-nonnegative}
\lim_{t\downarrow 0}\frac{D_{\alpha}(P_t\|Q) - D_{\alpha}(P_0\|Q)}{t} \ge 0.
\end{eqnarray}

\par
By Proposition~\ref{prop:pythagorean-property}a), if $\alpha \in (1, \infty)$,
$\text{Supp}(P^*) = \text{Supp}(\mathscr{L}_{\alpha})$.
Hence, if $\alpha > 1$, for every $P\in \mathscr{L}_{\alpha}$ there exists $t'<0$
such that $$(1-t)P^*(a)^{\alpha} + t P(a)^{\alpha} > 0$$ for all
$a\in\text{Supp}(\mathscr{L}_{\alpha})$ and $t\in (t',0)$. Since
$\mathcal{A}$ is finite, the derivative of $D_{\alpha}(P_t\|Q)$ exists at $t=0$.
In view of \eqref{eqn:alpha-linear-family} and since $P,P^*\in\mathscr{L}_{\alpha}$,
for every $t\in (t',0)$, there exist
$\theta_1^{(t)},\dots,\theta_k^{(t)}\in \mathbb{R}$ such that
\begin{align*}
(1-t)P^*(a)^{\alpha} + t P(a)^{\alpha} = \sum\limits_{i=1}^k \theta_i^{(t)} f_i(a)
\end{align*}
which yields that $P_t \in \mathscr{L}_{\alpha}$ for $t\in(t',0)$ (see \eqref{eqn:s-t}).
Consequently, since \eqref{eqn:nonnegative} also holds for all $t\in (t',0)$, then
\begin{eqnarray}
\label{eqn:derivative-nonpositive}
\lim_{t\uparrow 0}\frac{D_{\alpha}(P_t\|Q) - D_{\alpha}(P_0\|Q)}{t} \leq 0.
\end{eqnarray}
From \eqref{eqn:derivative-nonnegative}, \eqref{eqn:derivative-nonpositive}, and the existence
of the derivative of $D_{\alpha}(P_t\|Q)$ at $t=0$, it follows that this derivative should be
equal to zero; since this derivative is
equal to the right side of \eqref{eqn:derivative-calculation}, it follows that
\eqref{eqn:derivative-calculation} holds with equality. Hence, for every $P \in \mathcal{P}$,
\begin{align}
\label{eqn:pythagorean-expanded}
\frac{\sum_a P(a)^{\alpha}Q(a)^{1-\alpha}}{\sum_a P^*(a)^{\alpha}Q(a)^{1-\alpha}}
= \sum_a P(a)^{\alpha}P^*(a)^{1-\alpha}.
\end{align}
Taking logarithms on both sides of \eqref{eqn:pythagorean-expanded}, and dividing by
$\alpha-1$, yields \eqref{eqn:pythagorean-equality}.
\end{IEEEproof}
\vspace{0.1cm}

The following theorem suggests an iterative algorithm to find the forward
$D_{\alpha}$-projection when the underlying $\alpha$-convex set is an
intersection of a finite number of $\alpha$-linear families.

\begin{theorem}[Iterative projections]
\label{thm:iterative-projection}
Let $\alpha\in (1,\infty)$. Suppose that
$\mathscr{L}_{\alpha}^{(1)}, \dots, \mathscr{L}_{\alpha}^{(m)}$
are $\alpha$-linear families, and let
\begin{align}
\label{eqn:intersection-alpha-linear-families}
\mathcal{P} := \bigcap_{n=1}^m \mathscr{L}_{\alpha}^{(n)}
\end{align}
where $\mathcal{P}$ is assumed to be a non-empty set. Let $P_0 = Q$,
and let $P_n$ be the forward $D_{\alpha}$-projection of $P_{n-1}$ on
$\mathscr{L}_{\alpha}^{(i_n)}$ with $i_n =n \, \mathrm{mod} \, (m)$
for $n \in \naturals$. Then, $P_n \to P^*$ (a pointwise convergence
by letting $n \to \infty$).
\end{theorem}

\begin{IEEEproof}
Since (by definition) $P_n$ is a forward $D_{\alpha}$-projection of $P_{n-1}$
on an $\alpha$-linear set which includes $\mathcal{P}$ (see
\eqref{eqn:intersection-alpha-linear-families}), it follows from
Theorem~\ref{thm:pythagorean-equality} that for every $P \in \mathcal{P}$ and
$N \in \naturals$
\begin{align}
\label{eqn:pythagorean-for-iteration}
D_{\alpha}(P\|P_{n-1}) = D_{\alpha}(P\|P_n) + D_{\alpha}(P_n\|P_{n-1}),
\quad \forall \, n \in \{1,\dots,N\}.
\end{align}
Hence, since $P_0 = Q$, \eqref{eqn:pythagorean-for-iteration} yields
\begin{align}
\label{eqn:pythagorean-summed}
D_{\alpha}(P\|Q) &= D_{\alpha}(P\|P_N)
+ \sum_{n=1}^N \Bigl(D_{\alpha}(P\|P_{n-1}) - D_{\alpha}(P\|P_n)\Bigr) \nonumber \\
&= D_{\alpha}(P\|P_N) + \sum_{n=1}^N D_{\alpha}(P_n\|P_{n-1}).
\end{align}
Note that $\mathcal{P}$ in \eqref{eqn:intersection-alpha-linear-families}, being a non-empty
intersection of a finite number of compact sets, is a compact set. Let $\{P_{N_k}\}$
be a subsequence of $\{P_n\}$ in $\mathcal{P}$ which pointwise converges to some $P'$
on the finite set $\mathcal{A}$ (hence, $P' \in \mathcal{P}$). Letting $N_k \to \infty$
in \eqref{eqn:pythagorean-summed} implies that, for every $P \in \mathcal{P}$,
\begin{align}
\label{eqn:pythagorean-summed-infinity}
D_{\alpha}(P\|Q) = D_{\alpha}(P\|P') +
\sum_{n=1}^{\infty} D_{\alpha}(P_n\|P_{n-1})
\end{align}
where, to obtain \eqref{eqn:pythagorean-summed-infinity},
$D_{\alpha}(P\|P_{N_k}) \to D_{\alpha}(P\|P')$ since
$\mathcal{A}$ is finite and $P_{N_k} \to P'$.
Since \eqref{eqn:pythagorean-summed-infinity} yields
$\sum_{n=1}^{\infty} D_{\alpha}(P_n\|P_{n-1}) < \infty$ then
$D_{\alpha}(P_n\|P_{n-1}) \to 0$ as $n \to \infty$;
consequently, since $D_{\alpha}(\cdot \| \cdot)$ is monotonically
non-decreasing in $\alpha$ (see, e.g., \cite[Theorem~3]{ErvenH14})
and $\alpha > 1$ then $D(P_n\|P_{n-1}) \to 0$,
and by Pinsker's inequality $|P_n-P_{n-1}| \to 0$ as $n \to \infty$.
From the periodic construction of $\{i_n\}$ (with period $m$),
the subsequences $\{P_{N_k}\}, \{P_{N_k+1}\},\dots,\{P_{N_k+m-1}\}$ have
their limits in $\mathscr{L}_{\alpha}^{(1)},\dots,\mathscr{L}_{\alpha}^{(m)}$,
respectively. Since $|P_n-P_{n-1}| \to 0$ as $n\to\infty$, all these
subsequences have the same limit $P'$, which therefore implies that
$P' \in \mathcal{P}$.
Substituting $P=P'$ in \eqref{eqn:pythagorean-summed-infinity} yields
\begin{align} \label{eqn 2:pythagorean-summed-infinity}
D_{\alpha}(P'\|Q) = \sum_{n=1}^{\infty} D_{\alpha}(P_n\|P_{n-1})
\end{align}
and assembling \eqref{eqn:pythagorean-summed-infinity} and
\eqref{eqn 2:pythagorean-summed-infinity} yields
\begin{align} \label{eqn 3:pythagorean}
D_{\alpha}(P\|Q) = D_{\alpha}(P\|P') + D_{\alpha}(P'\|Q),
\quad \forall \, P \in \mathcal{P}.
\end{align}
Hence, \eqref{eqn 3:pythagorean} implies that $P'$ is the
forward $D_{\alpha}$-projection of $Q$ on $\mathcal{P}$. Since
$\{P_{N_k}\}$ is an arbitrary convergent subsequence of $\{P_n\}$,
and the forward $D_{\alpha}$-projection is unique, every convergent
subsequence of $\{P_n\}$ has the same limit $P^*$. This proves that
$P_n \to P^*$ as $n \to \infty$.
\end{IEEEproof}

\section{Forward Projection on an $\alpha$-Linear Family}
\label{sec:ForProjAlphaLinFam}

We identify in this section a parametric form of the forward
$D_{\alpha}$-projection on an $\alpha$-linear family, which turns out
to be a generalization of the well-known exponential family.

\vspace{0.1cm}
\begin{theorem}[Forward projection on an $\alpha$-linear family]
\label{thm:forward-projection}
Let $\alpha\in (0, 1)\cup (1, \infty)$, and let $P^*$ be the forward
$D_{\alpha}$-projection of $Q$ on an $\alpha$-linear family
$\mathscr{L}_{\alpha}$ (as defined in \eqref{eqn:alpha-linear-family}
where $f_1,\dots ,f_k$, as $|\mathcal{A}|$-dimensional vectors, are
mutually orthogonal). The following hold:
\begin{enumerate}[a)]
\item If $\text{Supp}(P^*) = \text{Supp}(\mathscr{L}_{\alpha})$, then
$P^*$ satisfies \eqref{eqn:pythagorean-equality}.
\item If
\begin{align}
\label{eqn:equal-supports}
\text{Supp}(P^*) = \text{Supp}(\mathscr{L}_{\alpha}) = \mathcal{A},
\end{align}
then there exist $f_{k+1},\dots, f_{|\mathcal{A}|}$ such that $f_1, \ldots,
f_{|\mathcal{A}|}$ are mutually orthogonal as $|\mathcal{A}|$-dimensional
vectors, and $\theta^* = (\theta_{k+1}^*,\dots ,\theta_{|\mathcal{A}|}^*)\in
\mathbb{R}^{|\mathcal{A}|-k}$ such that for all $a\in\mathcal{A}$
\vspace*{-0.3cm}
\begin{align}
\label{eqn:forward-projection}
& P^*(a) = Z(\theta^*)^{-1} \biggl[Q(a)^{1-\alpha} + (1-\alpha)
\sum\limits_{i=k+1}^{|\mathcal{A}|} \theta_i^* f_i(a)\biggr]^{\frac{1}{1-\alpha}}
\end{align}
where $Z(\theta^*)$ is a normalizing constant in \eqref{eqn:forward-projection}.
\end{enumerate}
\end{theorem}

\vspace{0.1cm}
\begin{IEEEproof}
The proof of Item~a) follows from the proof of
Theorem~\ref{thm:pythagorean-equality} which yields that $P^*$ satisfies
the Pythagorean equality \eqref{eqn:pythagorean-equality}.

We next prove Item~b). Eq.~\eqref{eqn:pythagorean-equality}
is equivalent to \eqref{eqn:pythagorean-expanded}, which can be re-written as
\begin{eqnarray}
\label{eqn:pythagorean-innerproduct}
\sum\limits_a P(a)^{\alpha} \left[c P^*(a)^{1-\alpha} - Q(a)^{1-\alpha}\right]
= 0, \quad \forall \, P \in \mathscr{L}_{\alpha}
\end{eqnarray}
with $c = \sum_a P^*(a)^{\alpha} Q(a)^{1-\alpha}$.
Recall that if a subspace of the Euclidean space $\mathbb{R}^{|\mathcal{A}|}$
contains a vector whose all components are strictly positive, then this
subspace is spanned by the vectors whose all components are nonnegative.
In view of \eqref{eqn:alpha-linear-family}, the subspace $\mathcal{F}$
which is spanned by $f_1, \ldots, f_k$ (recall that these functions are
regarded as $|\mathcal{A}|$-dimensional vectors) contains $(P^*)^{\alpha}$
whose support is $\mathcal{A}$ (see \eqref{eqn:equal-supports}).
Consequently, it follows from \eqref{eqn:alpha-linear-family}
that $\{P^{\alpha}\colon P\in\mathscr{L}_{\alpha}\}$
spans the subspace $\mathcal{F}$ of $\mathbb{R}^{|\mathcal{A}|}$.
From \eqref{eqn:pythagorean-innerproduct}, it also follows that
$c \; (P^*)^{1-\alpha} - Q^{1-\alpha} \in \mathcal{F}^{\perp}$, which
yields the existence of $\theta_i^* \in \mathbb{R}$ for
$i \in \{k+1,\dots, |\mathcal{A}|\}$ such that for all $a\in\mathcal{A}$
\begin{align} \label{haifa}
c P^*(a)^{1-\alpha} - Q(a)^{1-\alpha} = (1-\alpha) \,
\sum\limits_{i = k+1}^{|\mathcal{A}|} \theta_i^* f_i(a)
\end{align}
with a scaling of $\{\theta_i^*\}$ by $1-\alpha \neq 0$ in the right side
of \eqref{haifa}. Hence, $P^*$ satisfies \eqref{eqn:forward-projection}
where $c$ in the left side of \eqref{haifa} is the normalizing constant
$Z(\theta^*)$ in \eqref{eqn:forward-projection}.
\end{IEEEproof}

\begin{remark}
In view of Example~\ref{eg:counterexample}, the condition
$\text{Supp}(P^*) = \text{Supp}(\mathscr{L}_{\alpha})$ is not
necessarily satisfied for $\alpha\in(0,1)$. However, due to
Proposition~\ref{prop:pythagorean-property}~a), this condition
is necessarily satisfied for all $\alpha\in (1,\infty)$.
\end{remark}

\vspace*{0.2cm}
For $\alpha\in (0,\infty)$, the forward $D_{\alpha}$-projection on an $\alpha$-linear family
$\mathscr{L}_{\alpha}$ motivates the definition of the following parametric family of probability
measures. Let $Q\in \mathcal{M}$, and let
\begin{eqnarray}
\label{eqn:alpha-exponential-family}
\mathscr{E}_{\alpha} := \biggl\{P\in\mathcal{M}\colon P(a)
= Z(\theta)^{-1}\Biggl[Q(a)^{1-\alpha} + (1-\alpha)
\sum\limits_{i=k+1}^{|\mathcal{A}|} \theta_i f_i(a)\Biggr]^{\frac{1}{1-\alpha}}, \nonumber\\
\theta = (\theta_{k+1},\dots ,\theta_{|\mathcal{A}|})\in \mathbb{R}^{|\mathcal{A}|-k} \biggr\}.
\end{eqnarray}
We shall call the family $\mathscr{E}_{\alpha}$
an {\em $\alpha$-exponential family},\footnote{Note that the {\em $\alpha$-power-law family}
in \cite[Definition~8]{KumarS15b} is a different extension of the exponential family $\mathscr{E}$.}
which can be verified to be a $(1-\alpha)$-convex set.
We next show that $\mathscr{E}_{\alpha}$ generalizes the {\em exponential family} $\mathscr{E}$
defined in \cite[p.~24]{CsiszarS_FnT}:
\begin{eqnarray}
\label{eqn:exponential-family}
\mathscr{E} = \Biggl\{P\in\mathcal{M}\colon P(a) = Z(\theta)^{-1}Q(a)
\exp \Biggl( \, \sum\limits_{i=k+1}^{|\mathcal{A}|} \theta_i f_i(a) \Biggr), \nonumber \\
\theta = (\theta_{k+1},\dots ,\theta_{|\mathcal{A}|})\in \mathbb{R}^{|\mathcal{A}|-k}\Biggr\}.
\end{eqnarray}
To this end, let the {\em $\alpha$-exponential} and {\em $\alpha$-logarithm}
functions be, respectively, defined by
\begin{align}
\label{eq: e_alpha function}
 e_{\alpha}(x) & := \begin{cases}
\exp(x) & \mbox{if } \alpha = 1, \\[0.1cm]
\Bigl(\max\big\{1+(1-\alpha)x, \, 0\big\}\Bigr)^{\frac{1}{1-\alpha}}
& \mbox{if } \alpha \in (0,1)\cup (1,\infty), \\
\end{cases} \\[0.2cm]
\label{eq: ln alpha function}
\ln_{\alpha}(x) & := \begin{cases}
\ln(x) & \mbox{if } \alpha = 1, \\[0.1cm]
\frac{x^{1-\alpha}-1}{1-\alpha}
& \mbox{if } \alpha \in (0,1)\cup (1,\infty). \\
\end{cases}
\end{align}
In view of \eqref{eqn:alpha-exponential-family}, \eqref{eq: e_alpha function}
and \eqref{eq: ln alpha function}, the $\alpha$-exponential family
$\mathscr{E}_{\alpha}$ includes all the probability measures $P$
defined on $\mathcal{A}$ such that for all $a \in \mathcal{A}$
\begin{align}
\label{eq1: 26.11.15}
P(a) 
& = Z(\theta)^{-1} \,e_{\alpha} \Biggl(\ln_{\alpha}(Q(a))
+ \sum\limits_{i=k+1}^{|\mathcal{A}|} \theta_i f_i(a)\Biggr),
\end{align}
whereas any $P\in\mathscr{E}$ can be written as
\begin{align}
\label{eq3: 26.11.15}
P(a) = Z(\theta)^{-1} \exp \Biggl(\ln(Q(a)) +
\sum\limits_{i=k+1}^{|\mathcal{A}|} \theta_i f_i(a) \Biggr).
\end{align}
This is an alternative way to notice that the family $\mathscr{E}_{\alpha}$
can be regarded as a continuous extension of the exponential family
$\mathscr{E}$ when $\alpha\in (0,1)\cup (1,\infty)$.
It is easy to see that the reference measure $Q$ in the definition of
$\mathscr{E}_{\alpha}$ is always a member of $\mathscr{E}_{\alpha}$. As
in the case of the exponential family, the $\alpha$-exponential family
$\mathscr{E}_{\alpha}$ also depends on the reference measure $Q$ only in
a loose manner. In view of \eqref{eqn:alpha-exponential-family}, any other
member of $\mathscr{E}_{\alpha}$ can play the role of $Q$ in defining this
family. The proof is very similar to the one for the $\alpha$-power-law
family in \cite[Proposition~22]{KumarS15b}. It should be also noted that,
for $\alpha\in (1,\infty)$, all members of $\mathscr{E}_{\alpha}$ have the
same support (i.e., the support of $Q$).

\section{Orthogonality of $\alpha$-Linear and $\alpha$-Exponential Families}
\label{sec:Orthogonality}

In this section, we first prove an ``orthogonality" relationship between
an $\alpha$-exponential family and its associated $\alpha$-linear family.
We then use it to transform the reverse $D_{\alpha}$-projection on an
$\alpha$-exponential family into a forward $D_{\alpha}$-projection on
an $\alpha$-linear family.

Let us begin by making precise the notion of orthogonality between two
sets of probability measures with respect to $D_{\alpha}$ ($\alpha >0$).
\begin{definition}[Orthogonality of sets of probability measures]
Let $\alpha\in (0,1)\cup (1,\infty)$, and let $\mathcal{P}$ and $\mathcal{Q}$
be sets of probability measures. We say that {\em $\mathcal{P}$ is
$\alpha$-orthogonal to $\mathcal{Q}$ at  $P^*$} if the following hold:
\begin{itemize}
\item[1)] $\mathcal{P} \cap \mathcal{Q} = \{P^*\}$
\item[2)] $\label{pythagorean-inequality2}
D_{\alpha}(P\|Q) = D_{\alpha}(P\|P^*)+D_{\alpha}(P^*\|Q)$
for every $P \in \mathcal{P}$ and $Q \in \mathcal{Q}$.
\end{itemize}
\end{definition}

Note that, when $\alpha =1$, this refers to the orthogonality between
the linear and exponential families in \cite[Corollary~3.1]{CsiszarS_FnT}.

We are now ready to state our second main result namely, the orthogonality
between $\mathscr{L}_{\alpha}$ and $\mathscr{E}_{\alpha}$.
\vspace{0.2cm}

\begin{theorem}[Orthogonality of $\mathscr{L}_{\alpha}$ and $\mathscr{E}_{\alpha}$]
\label{thm:orthogonality}
Let $\alpha\in (1,\infty)$, let $\mathscr{L}_{\alpha}$ and
$\mathscr{E}_{\alpha}$ be given in \eqref{eqn:alpha-linear-family}
and \eqref{eqn:alpha-exponential-family}, respectively, and let
$P^*$ be the forward $D_{\alpha}$-projection of $Q$ on
$\mathscr{L}_{\alpha}$. The following hold:
\begin{enumerate}[a)]
\item $\mathscr{L}_{\alpha}$ is $\alpha$-orthogonal to
$\text{cl}(\mathscr{E}_{\alpha})$ at $P^*$.
\item If $\text{Supp}(\mathscr{L}_{\alpha}) = \mathcal{A}$,
then $\mathscr{L}_{\alpha}$ is $\alpha$-orthogonal to
$\mathscr{E}_{\alpha}$ at $P^*$.
\end{enumerate}
\end{theorem}

\begin{IEEEproof}
In view of Proposition~\ref{prop:pythagorean-property}~a), for $\alpha \in (1, \infty)$, the
condition $\text{Supp}(P^*) = \text{Supp}(\mathscr{L}_{\alpha})$ is satisfied. Consequently,
for $\alpha \in (1, \infty)$, Theorem~\ref{thm:forward-projection}a) implies that $P^*$
satisfies \eqref{eqn:pythagorean-equality}. We next prove the following:
\begin{itemize}
\item[i)] Every $\tilde{P}\in\mathscr{L}_{\alpha}\cap\text{cl}(\mathscr{E}_{\alpha})$
satisfies \eqref{eqn:pythagorean-equality} with $\tilde P$ in place of $P^*$.
\item[ii)] $\mathscr{L}_{\alpha}\cap\text{cl}(\mathscr{E}_{\alpha})$ is non-empty.
\end{itemize}
To prove Item~i), since $\tilde P\in\text{cl}(\mathscr{E}_{\alpha})$, there exists a
sequence $\{P_n\}$ in $\mathscr{E}_{\alpha}$ such that $P_n\to\tilde P$. Since $P_n\in\mathscr{E}_{\alpha}$,
from \eqref{eqn:alpha-exponential-family}, there exists
$\theta^{(n)} := (\theta_{k+1}^{(n)},\dots,\theta_{|\mathcal{A}|}^{(n)})\in\mathbb{R}^{|\mathcal{A}|-k}$
such that for all $a \in \mathcal{A}$
\begin{align}
\label{eqn:1}
P_n(a)^{1-\alpha} =  Z(\theta^{(n)})^{\alpha -1}\Big[Q(a)^{1-\alpha} +
(1-\alpha)\sum\limits_{i=k+1}^{|\mathcal{A}|}\theta_i^{(n)} f_i(a)\Big].
\end{align}
Since $P, \tilde{P} \in \mathscr{L}_{\alpha}$, from \eqref{eqn:linear-family},
for all $i \in \{k+1,\dots, |\mathcal{A}|\}$
\begin{align}
\label{eqn:2}
& \sum_a P(a)^{\alpha} f_i(a) = 0, \\
\label{eqn:3}
& \sum_a \tilde{P}(a)^{\alpha} f_i(a) = 0.
\end{align}
Since $\mathcal{A}$ is finite, assembling \eqref{eqn:1}--\eqref{eqn:3} yields
(after switching the order of summations over $a \in \mathcal{A}$ and
$i  \in \{k+1,\dots, |\mathcal{A}|\}$)
\begin{align}
\label{eqn:4}
& \sum_a  P(a)^{\alpha}P_n(a)^{1-\alpha} =  Z(\theta^{(n)})^{\alpha -1}\sum_a
P(a)^{\alpha}Q(a)^{1-\alpha}, \\
\label{eqn:5}
& \sum_a  \tilde{P}(a)^{\alpha}P_n(a)^{1-\alpha} =  Z(\theta^{(n)})^{\alpha -1}
\sum_a  \tilde{P}(a)^{\alpha}Q(a)^{1-\alpha},
\end{align}
and, from \eqref{eqn:4} and \eqref{eqn:5},
\begin{align}
\label{eqn:6}
\sum_a \tilde P(a)^{\alpha}P_n(a)^{1-\alpha} = \frac{\sum_a  P(a)^{\alpha}
P_n(a)^{1-\alpha}}{\sum_a  P(a)^{\alpha}Q(a)^{1-\alpha}}\cdot\sum_a
\tilde{P}(a)^{\alpha}Q(a)^{1-\alpha}.
\end{align}
Since $P_n\to \tilde P$, letting $n \to \infty$ in \eqref{eqn:6} yields
\begin{align}
\label{eqn:7}
1 = \frac{\sum_a  P(a)^{\alpha}\tilde P(a)^{1-\alpha}}{\sum_a
P(a)^{\alpha}Q(a)^{1-\alpha}}\cdot\sum_a \tilde{P}(a)^{\alpha}Q(a)^{1-\alpha},
\end{align}
which is equivalent to \eqref{eqn:pythagorean-equality} when $P^*$ is replaced by $\tilde P$.

\par
To prove Item~ii), note that if $\text{Supp}(\mathscr{L}_{\alpha}) = \mathcal{A}$, then
Theorem~\ref{thm:forward-projection}b) yields that $P^*\in\mathscr{L}_{\alpha}\cap
\mathscr{E}_{\alpha}$, and we are done. So suppose that $\text{Supp}(\mathscr{L}_{\alpha})
\neq \mathcal{A}$, and consider the following sequence of $\alpha$-linear families:
\begin{align}
\label{eqn:sequence-alpha-linear-family}
\mathscr{L}_{\alpha}^{(n)} := \Big\{P\in\mathcal{M}\colon\sum_a P(a)^{\alpha}\tilde f_i(a) = 0,
\quad i\in\{k+1,\dots,|\mathcal{A}|\}\Big\},
\end{align}
where
\begin{align} \label{eq: the rest of f_i's}
\tilde f_i(a) := f_i(a) - \eta_i^{(n)}Q(a)^{1-\alpha}, \quad \forall \, a \in \mathcal{A}
\end{align}
with
\begin{align} \label{eq: the rest of eta_i's}
\eta_i^{(n)} := \frac{\frac{1}{n}\sum_a Q(a)^{\alpha}f_i(a)}{(1-\frac{1}{n})
\sum_a P^*(a)^{\alpha}Q(a)^{1-\alpha} + \frac{1}{n}}, \quad i\in\{k+1,\dots,|\mathcal{A}|\}.
\end{align}
The $\tilde f_i$'s and $\eta_i^{(n)}$'s in \eqref{eq: the rest of f_i's} and \eqref{eq: the rest of eta_i's}
are selected such that the $\bigl(\alpha,\frac1{n}\bigr)$-mixture of $(P^*, Q)$ is a member of
$\mathscr{L}_{\alpha}^{(n)}$. This implies that $\text{Supp}(\mathscr{L}_{\alpha}^{(n)}) = \mathcal{A}$
(recall that we assume that $\text{Supp}(Q) = \mathcal{A}$).
Notice also that $\eta_i^{(n)}\to 0$ as $n\to\infty$. Hence, $\mathscr{L}_{\alpha}^{(n)}$ asymptotically
coincides with $\mathscr{L}_{\alpha}$ as $n\to\infty$. Now, let $P_n$ be the forward $D_{\alpha}$-projection
of $Q$ on $\mathscr{L}_{\alpha}^{(n)}$. Then by Proposition~\ref{prop:pythagorean-property},
$\text{Supp}(P_n) = \mathcal{A}$, and hence by Theorem~\ref{thm:forward-projection}, there exists
$\theta^{(n)} := (\theta_{k+1}^{(n)},\dots,\theta_{|\mathcal{A}|}^{(n)})\in\mathbb{R}^{|\mathcal{A}|-k}$
such that for all $a \in \mathcal{A}$
\begin{align}
\label{eqn:p-n-a}
P_n(a)^{1-\alpha}
& = Z(\theta^{(n)})^{\alpha -1}\Big[Q(a)^{1-\alpha} +
(1-\alpha)\sum\limits_{i=k+1}^{|\mathcal{A}|}\theta_i^{(n)} \tilde f_i(a)\Big]\\
\label{eqn:p-n-b}
& = Z(\theta^{(n)})^{\alpha -1}\Big[Q(a)^{1-\alpha} +
(1-\alpha)\sum\limits_{i=k+1}^{|\mathcal{A}|}\theta_i^{(n)} \bigl(f_i(a)
- \eta_i^{(n)} Q(a)^{1-\alpha}\bigr)\Big]\\
& = Z(\theta^{(n)})^{\alpha -1}\Big[\Big(1-(1-\alpha)
\sum\limits_{i=k+1}^{|\mathcal{A}|}\theta_i^{(n)}\eta_i^{(n)}\Big)Q(a)^{1-\alpha} \nonumber \\
\label{eqn:p-n-c}
& \hspace*{2.7cm} + (1-\alpha)\sum\limits_{i=k+1}^{|\mathcal{A}|}\theta_i^{(n)} f_i(a)\Big].
\end{align}
Multiplying the left side of \eqref{eqn:p-n-a} and the right side of \eqref{eqn:p-n-c} by $P^*(a)^{\alpha}$,
summing over all $a\in \mathcal{A}$, and using the fact that $\sum_a P^*(a)^{\alpha} f_i(a) = 0$ for all
$i \in \{k+1, \ldots, |\mathcal{A}|\}$ yields
\begin{align}
\sum_a P^*(a)^{\alpha}P_n(a)^{1-\alpha}
= Z(\theta^{(n)})^{\alpha -1}\Big(1-(1-\alpha)
\sum\limits_{i=k+1}^{|\mathcal{A}|}\theta_i^{(n)}
\eta_i^{(n)}\Big)\sum_a P^*(a)^{\alpha}Q(a)^{1-\alpha}.
\end{align}
This implies that the term
$1-(1-\alpha)\sum\limits_{i=k+1}^{|\mathcal{A}|}\theta_i^{(n)}\eta_i^{(n)}$
is positive for all $n$; hence, dividing the left side of \eqref{eqn:p-n-a}
and the right side of \eqref{eqn:p-n-c} by this positive term yields that
$P_n \in \mathscr{E}_{\alpha}$. This implies that the limit of every convergent
subsequence of $\{P_n\}$ is a member of $\text{cl}(\mathscr{E}_{\alpha})$,
as well as of $\mathscr{L}_{\alpha}$.

\par
In view of Items~i) and~ii), as listed at the beginning of this proof, it now
follows from Proposition~\ref{prop:pythagorean-property}~b) and
Corollary~\ref{corollary: uniqueness} that
$\mathscr{L}_{\alpha} \cap \text{cl}(\mathscr{E}_{\alpha}) = \{P^*\}$.
Recall that, for $\alpha \in (1, \infty)$, Theorem~\ref{thm:forward-projection}a)
implies that $P^*$ satisfies \eqref{eqn:pythagorean-equality}; furthermore,
since $Q$ in \eqref{eqn:alpha-exponential-family} can be replaced
by any other member of $\mathscr{E}_{\alpha}$, the satisfiability of
\eqref{eqn:pythagorean-equality} for $Q \in \mathscr{E}_{\alpha}$ yields
its satisfiability with any other member of $\mathscr{E}_{\alpha}$ replacing $Q$.
Since $\mathcal{A}$ is finite, \eqref{eqn:pythagorean-equality} is also satisfied
with any member of $\text{cl}(\mathscr{E}_{\alpha})$ replacing $Q$;
this can be justified for any $\widetilde{Q} \in \text{cl}(\mathscr{E}_{\alpha})$
by selecting a sequence $\{\widetilde{Q}_n\}$  in $\mathscr{E}_{\alpha}$
which pointwise converges to $\widetilde{Q}$, and by letting $n \to \infty$.
This proves Item~a).

\par
We next prove Item~b). Since by our assumption
$\text{Supp}(\mathscr{L}_{\alpha}) = \mathcal{A}$ and $\alpha \in (1, \infty)$
then Proposition~\ref{prop:pythagorean-property}~a) implies that
condition \eqref{eqn:equal-supports} holds.
From Proposition~\ref{prop:pythagorean-property}~b),
Corollary~\ref{corollary: uniqueness} and Theorem~\ref{thm:forward-projection},
it follows that the forward $D_{\alpha}$-projection $P^*$ is the unique member of
$\mathscr{L}_{\alpha}\cap \mathscr{E}_{\alpha}$
satisfying \eqref{eqn:pythagorean-equality}. Similarly to the previous paragraph,
\eqref{eqn:pythagorean-equality} is satisfied not only for for $Q \in \mathscr{E}_{\alpha}$,
but also for any other member of $\mathscr{E}_{\alpha}$ replacing $Q$.
This proves Item~b).
\end{IEEEproof}

\vspace*{0.1cm}
\begin{remark} \label{remark: orthogonality for alpha less than 1}
In view of Example \ref{eg:counterexample}, if $\alpha \in (0,1)$,
$\text{Supp}(P^*)$ is not necessarily equal to $\text{Supp}(\mathscr{L}_{\alpha})$;
this is consistent with Theorem~\ref{thm:orthogonality}
which is stated only for $\alpha\in (1,\infty)$.
Nevertheless, in view of the proof of Theorem~\ref{thm:pythagorean-equality},
the following holds for $\alpha\in (0,1)$: if the condition
$\text{Supp}(P^*) = \text{Supp}(\mathscr{L}_{\alpha}) = \mathcal{A}$ is satisfied,
then $\mathscr{L}_{\alpha}$ is $\alpha$-orthogonal to $\mathscr{E}_{\alpha}$ at $P^*$.
\end{remark}

\section{Reverse Projection on an $\alpha$-Exponential Family}
\label{sec:RevProjAlphaExpFam}

In this section, we define reverse $D_{\alpha}$-projections, and we rely on
the orthogonality property in Theorem~\ref{thm:orthogonality}
(and the note in Remark~\ref{remark: orthogonality for alpha less than 1})
to convert the reverse $D_{\alpha}$-projection on an $\alpha$-exponential
family into a forward projection on an $\alpha$-linear family.

\vspace{0.1cm}
\begin{definition}[Reverse $D_{\alpha}$-projection]
Let $P\in \mathcal{M}$, $\mathcal{Q}\subseteq\mathcal{M}$, and $\alpha \in (0, \infty)$.
If there exists $Q^*\in \mathcal{Q}$ which attains the global minimum of
$D_{\alpha}(P\|Q)$ over all $Q\in \mathcal{Q}$ and $D_{\alpha}(P\|Q^*) < \infty$,
then $Q^*$ is said to be a {\em reverse $D_{\alpha}$-projection} of $P$ on $\mathcal{Q}$.
\end{definition}
\vspace{0.1cm}

\begin{theorem}
\label{thm:application-orthogonality}
Let $\alpha \in (0,1) \cup (1,\infty)$, and let $\mathscr{E}_{\alpha}$ be an
$\alpha$-exponential family determined by $Q, f_{k+1},\dots,f_{|\mathcal{A}|}$.
Let $X_1,\dots,X_n$ be i.i.d. samples drawn at random according to a probability
measure in $\mathscr{E}_{\alpha}$. Let $\hat{P_n}$ be the empirical probability
measure of $X_1,\dots,X_n$, and let $P_n^*$ be the forward
$D_{\alpha}$-projection of $Q$ on the $\alpha$-linear family
\begin{align}
\label{eqn:empirical-alpha-linear-family}
\mathscr{L}_{\alpha}^{(n)} := \Big\{P\in\mathcal{M} \colon
\sum_{a \in \mathcal{A}} P(a)^{\alpha} \hat{f_i}(a) = 0, \quad i\in\{k+1,\dots,|\mathcal{A}|\}\Big\},
\end{align}
where
\begin{align}
\label{eq: shifted f_i}
\hat{f_i}(a) := f_i(a) - \hat{\eta}_i^{(n)} Q(a)^{1-\alpha}, \quad
\forall \, a\in\mathcal{A}
\end{align}
with
\begin{align}
\label{eq: eta_i's}
\hat{\eta}_i^{(n)} := \frac{\sum_a \hat{P}_n(a)^{\alpha} f_i(a)}{\sum_a
\hat{P}_n(a)^{\alpha}Q(a)^{1-\alpha}}, \quad i\in\{k+1,\dots,|\mathcal{A}|\}.
\end{align}
The following hold:
\begin{enumerate}[a)]
\item If $\text{Supp}(\mathscr{L}_{\alpha}^{(n)}) = \mathcal{A}$ for $\alpha \in (1, \infty)$
or $\text{Supp}(P_n^*) = \text{Supp}(\mathscr{L}_{\alpha}^{(n)}) = \mathcal{A}$ for $\alpha \in (0,1)$,
then $P_n^*$ is the reverse $D_{\alpha}$-projection of $\hat{P_n}$ on
$\mathscr{E}_{\alpha}$.

\item For $\alpha\in (1,\infty)$, if $\text{Supp}(\mathscr{L}_{\alpha}^{(n)})
\neq \mathcal{A}$, then the reverse $D_{\alpha}$-projection of $\hat{P}_n$ on
$\mathscr{E}_{\alpha}$ does not exist. Nevertheless, $P_n^*$ is the reverse
$D_{\alpha}$-projection of $\hat{P_n}$ on $\text{cl}(\mathscr{E}_{\alpha})$.
\end{enumerate}
\end{theorem}

\begin{IEEEproof}
To prove Item~a), note that $\mathscr{L}_{\alpha}^{(n)}$ in
\eqref{eqn:empirical-alpha-linear-family}--\eqref{eq: eta_i's}
is constructed in such a way that
\begin{align} \label{eq: P_n is a member}
\hat{P_n} \in \mathscr{L}_{\alpha}^{(n)}.
\end{align}
Following
\eqref{eqn:alpha-exponential-family}, let $\mathscr{E}_{\alpha} = \mathscr{E}_{\alpha}(f_{k+1},
\dots, f_{|\mathcal{A}|}; Q)$ denote the $\alpha$-exponential family determined
by $f_{k+1},\dots,f_{|\mathcal{A}|}$ and $Q$. We claim that
\begin{align} \label{eq: equality of exp. families}
\mathscr{E}_{\alpha}(f_{k+1},\dots,f_{|\mathcal{A}|};Q)
= \mathscr{E}_{\alpha}(\hat f_{k+1},\dots,\hat f_{|\mathcal{A}|};Q).
\end{align}
Indeed, if $P\in\mathscr{E}_{\alpha}(f_{k+1},\dots,f_{|\mathcal{A}|};Q)$,
then there exist $\theta = (\theta_{k+1},\dots ,\theta_{|\mathcal{A}|})\in \mathbb{R}^{|\mathcal{A}|-k}$
and a normalizing constant $Z = Z(\theta)$ such that for all $a \in \mathcal{A}$
\begin{align}
\label{eq1: p-intermsof-hatf}
P(a)^{1-\alpha} & = Z^{\alpha -1} \Big[Q(a)^{1-\alpha}
+ (1-\alpha)\sum_{i=k+1}^{|\mathcal{A}|}\theta_i f_i(a)\Big] \\
\label{eq2: p-intermsof-hatf}
& = Z^{\alpha -1} \Big[\Big(1+(1-\alpha)\sum_{i=k+1}^{|\mathcal{A}|}
\theta_i\hat{\eta}_i^{(n)}\Big)Q(a)^{1-\alpha}
+ (1-\alpha)\sum_{i=k+1}^{|\mathcal{A}|}\theta_i \hat f_i(a)\Big]
\end{align}
where \eqref{eq1: p-intermsof-hatf} and \eqref{eq2: p-intermsof-hatf}
follow, respectively, from \eqref{eqn:alpha-exponential-family} and
\eqref{eq: shifted f_i}.
Multiplying the left side of \eqref{eq1: p-intermsof-hatf} and the
right side of \eqref{eq2: p-intermsof-hatf} by $\hat{P_n}(a)^{\alpha}$,
summing over all $a\in\mathcal{A}$, and using \eqref{eq: P_n is a member} yields
\begin{align} \label{eq: haifa}
\sum_a \hat{P_n}(a)^{\alpha} P(a)^{1-\alpha} = Z^{\alpha -1}\Biggl(1+(1-\alpha)
\sum_{i=k+1}^{|\mathcal{A}|} \theta_i \hat{\eta}_i^{(n)}\Biggr)
\sum_a \hat{P_n}(a)^{\alpha} Q(a)^{1-\alpha}.
\end{align}
Eq.~\eqref{eq: haifa} yields
$1+(1-\alpha)\sum\limits_{i=k+1}^{|\mathcal{A}|}\theta_i\hat{\eta}_i^{(n)} > 0$.
Consequently, by rescaling \eqref{eq2: p-intermsof-hatf} appropriately,
it follows that
$P \in \mathscr{E}_{\alpha}(\hat f_{k+1},\dots,\hat f_{|\mathcal{A}|}; Q)$
which therefore implies that
\begin{align} \label{eq: indore}
\mathscr{E}_{\alpha}(f_{k+1}, \ldots, f_{|\mathcal{A}|}; Q)
\subseteq \mathscr{E}_{\alpha}(\hat f_{k+1},\dots,\hat f_{|\mathcal{A}|};Q).
\end{align}
Similarly, one can show that the reverse relation of \eqref{eq: indore} also
holds, which yields \eqref{eq: equality of exp. families}.
The proof of a) is completed by considering the following two cases:
\begin{itemize}
\item If $\alpha \in (1, \infty)$ and
$\text{Supp}(\mathscr{L}_{\alpha}^{(n)}) = \mathcal{A}$,
in view of Theorem~\ref{thm:orthogonality}b), $\mathscr{L}_{\alpha}^{(n)}$ is
$\alpha$-orthogonal to
$\mathscr{E}_{\alpha} = \mathscr{E}_{\alpha}(\hat f_{k+1},\dots,\hat f_{|\mathcal{A}|};Q)$
at $P_n^*$; hence, due to \eqref{eq: P_n is a member},
\begin{align}
\label{eqn:pythagorean-empirical}
D_{\alpha}(\hat{P_n}\|Q) = D_{\alpha}(\hat{P_n}\|P_n^*) +
D_{\alpha}(P_n^*\|Q), \quad \forall\,Q\in \mathscr{E}_{\alpha}.
\end{align}
Since $P_n^*\in\mathscr{E}_{\alpha}$, the minimum of
$D_{\alpha}(\hat{P_n}\|Q)$ subject to $Q\in\mathscr{E}_{\alpha}$ is uniquely
attained at $Q = P_n^*$.
\item If $\alpha \in (0,1)$ and
$\text{Supp}(P_n^*) = \text{Supp}(\mathscr{L}_{\alpha}^{(n)}) = \mathcal{A}$,
then \eqref{eqn:pythagorean-empirical} holds in view of Remark~\ref{remark: orthogonality for alpha less than 1}
and \eqref{eq: P_n is a member}. The minimum of $D_{\alpha}(\hat{P_n}\|Q)$
subject to $Q\in\mathscr{E}_{\alpha}$ is thus uniquely attained at $Q = P_n^*$.
\end{itemize}

To prove Item~b), for $\alpha\in (1,\infty)$, note that $P_n^*\in\mathscr{E}_{\alpha}$
if and only if $\text{Supp}(\mathscr{L}_{\alpha}^{(n)}) = \mathcal{A}$. Indeed,
the 'if' part follows from Item~a). The 'only if'
part follows from the fact that all members of $\mathscr{E}_{\alpha}$ have the same
support, $Q$ is a member of $\mathscr{E}_{\alpha}$ which by assumption has full support,
and $P_n^*$ is in both $\mathscr{E}_{\alpha}$ (by assumption) and $\mathscr{L}_{\alpha}^{(n)}$
(by definition).

To prove the first assertion in Item~b), note that by Theorem~\ref{thm:orthogonality}a),
$P_n^* \in \text{cl}(\mathscr{E}_{\alpha})$ and \eqref{eqn:pythagorean-empirical} holds
for every $Q \in \text{cl}(\mathscr{E}_{\alpha})$. Hence,
\begin{align}   \label{eq: min over closure of E_alpha}
\min_{Q \in \text{cl}(\mathscr{E}_{\alpha})} D_{\alpha}(\hat{P_n}\|Q) = D_{\alpha}(\hat{P_n} \| P_n^*).
\end{align}
Due to the continuity of $D_{\alpha}(\hat{P_n}\|Q)$ for $Q$ which is defined on the finite set
$\mathcal{A}$, it follows from \eqref{eq: min over closure of E_alpha} that
\begin{align}
\label{eq: inf over E alpha}
\inf_{Q \in \mathscr{E}_{\alpha}}  D_{\alpha}(\hat{P_n}\|Q)
& = D_{\alpha}(\hat{P_n} \| P_n^*).
\end{align}
In view of \eqref{eqn:pythagorean-empirical}, the minimum of
$D_{\alpha}(\hat{P_n}\|Q)$ over $Q \in \mathscr{E}_{\alpha}$ is not attained.
Finally, the last assertion in b) is due to \eqref{eqn:pythagorean-empirical} which, in view of
Theorem~\ref{thm:orthogonality}a), holds for all $Q\in\text{cl}(\mathscr{E}_{\alpha})$.
\end{IEEEproof}

\section{Summary and Concluding Remarks}
\label{sec:summary}

In \cite[Theorem~14]{ErvenH14}, van Erven and Harremo\"es proved a Pythagorean inequality
for R\'{e}nyi divergences on $\alpha$-convex sets under the assumption that the forward
projection exists. Motivated by their result, we study forward and reverse projections
for the R\'{e}nyi divergence of order $\alpha$ on $\alpha$-convex sets.
The results obtained in this paper, for $\alpha \in (0,\infty)$, generalize the known results for
$\alpha=1$; this special case corresponds to projections of the relative entropy on
convex sets, as studied by Csisz\'{a}r {\em et al.} in \cite{Csiszar75}, \cite{Csiszar84},
\cite{CsiszarM03}, \cite{CsiszarS_FnT}. The main contributions of this paper are as follows:
\begin{enumerate}[1)]
\item we prove a sufficient condition for the existence of a forward
projection in the general alphabet setting.
\item we prove a projection theorem on an $\alpha$-linear
family in the finite alphabet setting, and the parametric form
of this projection gives rise to an $\alpha$-exponential family.
\item we prove an orthogonality property between $\alpha$-linear and
$\alpha$-exponential families; it yields a duality between forward
and reverse projections, respectively, on these families.
\item we prove a convergence result of an iterative
algorithm for calculating the forward projection on
an intersection of a finite number of $\alpha$-linear families.
\end{enumerate}

For $\alpha=0$, the notion of an $\alpha$-convex set is continuously extended to a $\log$-convex
set. Since $D_0(P \| Q) = -\log Q\bigl(\text{Supp}(P)\bigr)$ (see, e.g., \cite[Theorem~4]{ErvenH14}),
if there exists $P \in \mathcal{P}$ such that $\text{Supp}(P) = \text{Supp}(Q)$ then any
such probability measure is a forward $D_0$-projection of $Q$ on $\mathcal{P}$ for which $D_0(P \| Q)=0$.
Note that, in this case, a forward $D_0$-projection of $Q$ on $\mathcal{P}$ is not necessarily unique.

For $\alpha=0$ and a finite set $\mathcal{A}$, the
notion of an $\alpha$-linear family is the whole simplex of probability measures
(with the convention that $0^0=1$ in \eqref{eqn:linear-family}),
provided that $\sum_a f_i(a)=0$ for all $i \in \{k+1, \ldots, |\mathcal{A}|\}$; otherwise,
the 0-linear family is an empty set. In the former case, the forward $D_0$-projection of $Q$ on $\mathcal{P}$
is any probability measure $P$ with a full support since in this case
$D_0(P \| Q)=0$; the forward $D_0$-projection is, however,
meaningless in the latter case where $\mathcal{P}$ is an empty set.

The R\'enyi divergence of order $\infty$ is well defined (see, e.g., \cite[Theorem~6]{ErvenH14});
furthermore, a set is defined to be $\infty$-convex if for all $P_0, P_1 \in \mathcal{P}$,
the probability measure $S_{0,1}$ whose $\mu$-density $s_{0,1}$ is equal to the
normalized version of $\max\{p_0, p_1\}$, is also included in $\mathcal{P}$ (this
definition follows from \eqref{eqn:alpha-lambda-mixture} by letting $\alpha \to \infty$). In
this case, Theorems~\ref{theorem: existence of forward Dalpha projection}
and~\ref{thm:pythagorean-equality} continue to hold for $\alpha =\infty$
(recall that Theorem~\ref{thm:pythagorean-equality} refers to the setting where
$\mathcal{A}$ is finite).

Consider the case where $\alpha = \infty$ and $\mathcal{A}$ is a finite set.
By continuous extension, the $\infty$-linear family necessarily includes all
the probability measures that are not a point mass (see \eqref{eqn:linear-family}),
and the $\infty$-exponential family only includes the reference measure $Q$
(see \eqref{eqn:alpha-exponential-family}). Consequently, the results in
Theorems~\ref{thm:iterative-projection}--\ref{thm:application-orthogonality}
become trivial for $\alpha = \infty$.

\section*{Acknowledgment}
This work has been supported by the Israeli Science Foundation (ISF) under Grant 12/12.
We thank the Associate editor, the anonymous reviewers of this journal paper and the ISIT~'16
conference paper for their valuable feedback which helped to improve the presentation.

\eject

\end{document}